\documentclass{aa}

\usepackage{euclid}
\usepackage{graphicx}
\usepackage{natbib}
\usepackage{scalerel}

\usepackage[table]{xcolor}

\usepackage{txfonts}
\usepackage[pdfencoding=auto,psdextra]{hyperref}
\hypersetup{
colorlinks=true,
linkcolor=blue,
filecolor=magenta,      
urlcolor=blue,
citecolor=blue
}
\makeatletter
\renewcommand*\aa@pageof{, page \thepage{} of \pageref*{LastPage}}
\makeatother

\usepackage[utf8]{inputenc}
\usepackage{subcaption}
\usepackage[switch, modulo]{lineno}

\usepackage{amsmath}	
\usepackage{dsfont}
\usepackage{newtxtext,newtxmath}
\usepackage{tikz}
\usetikzlibrary{angles, quotes}
\usetikzlibrary{shapes.arrows,calc}
\usepackage{physics} 
\usepackage[margin=10pt,font=small,labelfont=bf]{caption}

\newcommand{\Omm}{\Omega_\mathrm{m}}
\newcommand{\Map}{\mathcal{M}_\mathrm{ap}}

\newcommand{\MapMapMap}{{\expval{\Map^3}}}

\newcommand{\MapEst}{\widehat{\mathcal{M}}_\mathrm{ap}}

\renewcommand{\ee}{\mathrm{e}}

\newcommand{\ttheta}[1]{\pmb{\theta_\mathrm{ap}}}

\newcommand{\Norm}[2]{\mathcal{N}}

\defcitealias{Asgari2021}{A21}
\defcitealias{Hildebrandt2021}{H21}
\defcitealias{Takahashi2017}{T17}
\defcitealias{Heydenreich2023}{H23}

\tikzset{Pfeil/.style=%
{to path={let \p1 = ($(\tikztotarget)-(\tikztostart)$),
\n1 = {int(mod(scalar(atan2(\y1,\x1))+360, 360))}, 
\n2 = {veclen(\x1,\y1)}
in \pgfextra{\typeout{\n1,\n2,\x1,\y1}}
(\tikztotarget)
node[draw,single arrow,
minimum height=\n2-\pgflinewidth,
inner sep=1ex,
single arrow head extend=1ex,
rotate=\n1, 
anchor=tip, 
]{}
}}}

\begin{document}
%
%
\title{KiDS-1000 cosmology: Combined second- and third-order shear statistics}

\newcommand{\orcid}[1]{} 
\author{
Pierre A. Burger$^{1}$\thanks{E-mail: pburger@astro.uni-bonn.de},
Lucas Porth$^{1}$,
Sven Heydenreich$^{1,3}$,
Laila Linke$^{1,2}$,
Niek Wielders$^{1}$,
Peter Schneider$^{1}$,
Marika Asgari$^{4}$,
Tiago Castro$^{5,6,7}$,
Klaus Dolag$^{8,9}$,
Joachim Harnois-D\'eraps$^{10}$,
Hendrik Hildebrandt$^{11}$,
Konrad Kuijken$^{12}$
\and Nicolas Martinet$^{13}$
}
\institute{$^{1}$ University of Bonn, Argelander-Institut f\"ur Astronomie, Auf dem H\"ugel 71, 53121 Bonn, Germany\\
$^{2}$ Universit\"at Innsbruck, Institut f\"ur Astro- und Teilchenphysik, Technikerstr. 25/8, 6020 Innsbruck, Austria\\
$^{3}$ Department of Astronomy and Astrophysics, University of California, Santa Cruz, 1156 High Street, Santa Cruz, CA 95064 USA\\
$^{4}$ E.A Milne Centre, University of Hull, Cottingham Road, Hull, HU6 7RX, UK\\
$^{5}$  INAF -- Osservatorio Astronomico di Trieste, via Tiepolo 11, I-34131 Trieste, Italy\\
$^{6}$  INFN -- Sezione di Trieste, I-34100 Trieste, Italy\\
$^{7}$  IFPU -- Institute for Fundamental Physics of the Universe, via Beirut 2, 34151, Trieste, Italy\\
$^{8}$ Universitäts-Sternwarte, Fakultät für Physik, Ludwig-Maximilians-Universität München, Scheinerstr.1, 81679 München, Germany\\
$^{9}$ Max-Planck-Institut für Astrophysik, Karl-Schwarzschild-Straße 1, 85741 Garching, Germany\\
$^{10}$ School of Mathematics, Statistics and Physics, Newcastle University, Newcastle upon Tyne, NE1 7RU, UK \\
$^{11}$ Ruhr University Bochum, Faculty of Physics and Astronomy, Astronomical Institute (AIRUB), German Centre for Cosmological Lensing, 44780 Bochum, Germany \\
$^{12}$ Leiden Observatory, Leiden University, P.O.Box 9513, 2300RA Leiden, The Netherlands \\
$^{13}$ Aix-Marseille Université, CNRS, CNES, LAM, Marseille, France}

\date{Received 15 September 2023 / Accepted 6 December 2023}
%
%

\abstract
{}
{In this work, we perform the first cosmological parameter analysis of the 
fourth release of Kilo Degree Survey (KiDS-1000) data with second- and third-order shear statistics. This paper builds on a series of studies aimed at describing the roadmap to third-order shear statistics.}
{We derived and tested a combined model of the second-order shear statistic, namely, the COSEBIs and the third-order aperture mass statistics $\MapMapMap$ in a tomographic set-up. We validated our pipeline with $N$-body mock simulations of the KiDS-1000 data release. To model the second- and third-order statistics, we used the latest version of \textsc{HMcode2020} for the power spectrum and \textsc{BiHalofit} for the bispectrum. Furthermore, we used an analytic description to model intrinsic alignments and hydro-dynamical simulations to model the effect of baryonic feedback processes. Lastly, we decreased the dimension of the data vector significantly by considering only equal smoothing radii for the $\MapMapMap$ part of the data vector. This makes it possible to carry out a data analysis of the KiDS-1000 data release using a combined analysis of COSEBIs and third-order shear statistics.}
{We first validated the accuracy of our modelling by analysing a noise-free mock data vector, assuming the KiDS-1000 error budget, finding a shift in the maximum of the posterior distribution of the matter density parameter, $\Delta \Omm < 0.02\, \sigma_{\Omm}$, and of the structure growth parameter, $\Delta S_8 < 0.05\, \sigma_{S_8}$. Lastly, we performed the first KiDS-1000 cosmological analysis using a combined analysis of second- and third-order shear statistics, where we constrained $\Omm=0.248^{+0.062}_{-0.055}$ and $S_8=\sigma_8\sqrt{\Omm/0.3}=0.772\pm0.022$. The geometric average on the errors of $\Omm$ and $S_8$ of the combined statistics decreases, compared to the second-order statistic, by a factor of 2.2.}
{}
%
%
\keywords{gravitational lensing: weak – methods: numerical – large-scale structure of Universe}
%
%
\titlerunning{KiDS-1000 cosmology: Combined second- and third-order shear statistics}
\authorrunning{P. A. Burger et al.}
\maketitle
%
%
%
%

\section{Introduction}
Gravitational lensing describes the deflection of light by massive objects. It is sensitive to baryonic and dark matter and, therefore, ideal for probing the total matter distribution in the Universe. Since the distribution of matter is highly sensitive to cosmological parameters, it is excellent to test and probe the standard model of cosmology, called the $\Lambda$ Cold Dark Matter model ($\Lambda$CDM). Although the $\Lambda$CDM model can describe observations of the early Universe, such as the cosmic microwave background \citep[CMB; e.g.][]{planck2020}, or the Local Universe, such as the observed large-scale structure (LSS) of matter and galaxies \citep{Boss2017}, with remarkable accuracy, it is being put under stress due to tension observed between early and local probes. A $\sim 2 \,\sigma$ tension is in the structure growth parameter $S_8=\sigma_8\sqrt{\Omm/0.3}$, where $\sigma_8$ is the normalisation of the power spectrum and $\Omm$ is the total matter density parameter \citep[][]{Hildebrandt:2017, planck2020, joudaki2020, heymans2021,DES2021,diValentino:2021s,HSC2023}, suggests that the Local Universe is less clustered than what is expected from early-time measurements when extrapolated under the $\Lambda$CDM model. If these tensions are not due to systematics, extensions to the $\Lambda$CDM model are necessary, which will be tested with the next generation of cosmic shear surveys such as {\it Euclid} \citep{Laureijs:2011} or the Vera Rubin Observatory Legacy Survey of Space and Time \citep[LSST,][]{Ivezic:2008}. 

Commonly, two-point statistics for weak lensing and galaxy positions are used to infer cosmological parameters since they can be modelled accurately and systematic inaccuracies are well understood \citep{Schneider:1998,Troxel:2018,Hildebrandt:2017,Hikage:2019,Asgari:2019,Asgari:2020,Hildebrandt:2020}. Two-point statistics are excellent for capturing the entire information content of a Gaussian random field. However, non-linear gravitational instabilities created a significant amount of non-Gaussian features during the evolution of the Universe, such that the local matter distribution departed strongly from a Gaussian field. Therefore, higher order statistics are needed to extract all the available information in the local LSS of matter and galaxies. Furthermore, such higher order statistics usually depend differently on cosmological parameters and systematic effects such as intrinsic alignment (IA), meaning that a joint investigation of second- and higher order statistics tightens cosmological parameter constraints \citep[see, e.g.][]{Kilbinger:2005,Berge2010,Pires:2012,Fu:2014,Pyne:2021}. Recently used examples of higher order statistics are the peak count statistics \citep{Martinet:2018,Harnois-Deraps:2021}, persistent homology \citep{Heydenreich:2021}, density split statistics \citep{Gruen:2018,Burger2023}, and the integrated three-point correlation function used in \citep[][]{Halder:2021,Halder:2022}, along with a second- and third-order convergence moment analysis \citep{Gatti2022}. 

This work considers second- and third-order shear statistics, where the former probes the variance and the latter the skewness of the LSS at various scales. Our chosen second-order statistic is the $E_n$-modes of the COSEBIs \citep{Schneider2010}, and the third-order statistic $\MapMapMap$ is described in \cite{Schneider:2005} and recently measured in the Dark Energy Survey Year (DES) 3 Results \citep{Secco:2022}. Furthermore, this work belongs to a series of papers that aim for cosmological parameter analyses using third-order shear statistics. In the first paper,
\citet[][hereafter \citetalias{Heydenreich2023}]{Heydenreich2023}, we validated the analytical fitting formulae for a non-tomographic analysis and the conversion from three-point correlation function (3PCF) of cosmic shear to $\MapMapMap$. We found that $\MapMapMap$, even though they are combined from the shear 3PCF at different scales, contain a similar amount of information on $\Omm$ and $S_8$ as the 3PCF itself. The fact that $\MapMapMap$ can be measured from 3PCF is very convenient since this allows unbiased estimates for any survey geometry. A fast computation method of the aperture mass statistics by measuring the shear 3PCF is tackled in \cite{Porth2023}. Lastly, adding third-order statistics increases the dimension of the data vector significantly, and an analytical expression for the covariance is preferred, which is derived and validated for $\MapMapMap$ in a non-tomographic setup in \cite{Linke2023}. However, as this analysis considers combining second-order statistics with $\MapMapMap$ in a tomographic setup, and a joint covariance matrix has not been derived yet (Wielders et al. in prep), we must still rely on numerical simulations to determine the covariance matrix.
    
This article presents the first cosmological parameter analysis using the fourth data release of KiDS (KiDS-1000), combining second- and third-order shear statistics. 
We show the cosmological results, preceded by validation of several extensions of the analytical model to allow for a tomographic analysis and to include astrophysical effects such as IA \citep[e.g.][]{Joachimi2015} or baryonic feedback processes \citep{Chisari2015}. We aim to find the smallest set of scales that retain most of the cosmological information. 

The paper is structured as follows: In Sect.~\ref{sec:Theoretical background}, we review the basics of the second- and third-order shear statistics, extend the modelling to a tomographic analysis and describe our method to model the IA analytically. In Sect.~\ref{Sect:real_Data}, we describe the KiDS-1000 data and in Sect.~\ref{Sect:Data}, we introduce the simulation data used to validate our model and to correct it for baryonic feedback processes. In Sect.~\ref{sec:inference_description}, we briefly review our method to perform cosmological parameter interference. In Sect.~\ref{sec:combination_validation}, and \ref{sec:estimator_validation}, we validate our analysis pipeline against several systematics. The final cosmological results are presented in Sect.~\ref{sec:KiDS_results}, and we present our conclusions in Sect.~\ref{sec:Conclusions}.

\section{Theoretical background}

\label{sec:Theoretical background}
This section offers a review of the basics of weak gravitational lensing formalism and aperture statistics. For more detailed reviews, we refer to \citet{Bartelmann:2001}, \citet{Hoekstra:2008}, \citet{Munshi:2008}, \citet{Bartelmann:2010}, and \cite{Kilbinger2015}. In this work, we assume a spatially flat universe, such that the comoving angular-diameter distance is expressed as $f_{K}[\chi(z)]=\chi(z)$, where $\chi(z)$ is the comoving distance at redshift $z$. Given the matter density ,$\rho(\pmb{x},z)$, at comoving position, $\pmb{x}$, and redshift, $z$, the density contrast is $\delta(\pmb{x},z) = \frac{\rho(\pmb{x},z)}{\bar{\rho}(z)}-1$, where $\bar{\rho}(z)$ is the average matter density at redshift, $z$. The dimensionless surface mass density or convergence, $\kappa$, for sources at redshift, $z$, is determined by the line-of-sight integration as
\begin{align}
\kappa(\pmb{\vartheta},z) = \frac{3\Omm H_0^2}{2 c^2}\int_0^{\chi(z)}\dd \chi'{}&{}\, \frac{\chi'\,[\chi(z)-\chi']}{\chi(z)}
\frac{\delta(\chi'\pmb{\vartheta},z)}{a(\chi')}\; ,
\end{align}
where $\pmb{\vartheta}$ is the angular position on the sky, $H_0$ is the Hubble constant, and $a$ is the scale factor. The second argument of $\delta$ simultaneously describes the radial direction and the cosmological epoch, related through the light-cone condition $|c\, \dd t| = a(z) \, \dd \chi$.

\subsection{Limber projections of power- and bispectrum}
\label{subsec:power_and_bispectra_theory}
Given the Fourier transform $\hat{\delta}$ of the matter density contrast, the matter power spectrum, $P_{\delta\delta}(k,z)$, and bispectrum, $B_{\delta\delta\delta}(k_1,k_2,k_3,z)$, are:
\begin{align}
\expval{\hat{\delta}(\pmb{k}_1,z)\hat{\delta}(\pmb{k}_2,z)} = {}&{} (2\pi)^3\,\delta_\mathrm{D}(\pmb{k}_1+\pmb{k}_2)\,P_{\delta\delta}(k_1,z) \; , \\
\expval{\hat{\delta}(\pmb{k}_1,z)\hat{\delta}(\pmb{k}_2,z)\hat{\delta}(\pmb{k}_3,z)} = {}&{}  (2\pi)^3\,\delta_\mathrm{D}(\pmb{k}_1+\pmb{k}_2+\pmb{k}_3) \nonumber\\
{}&{}\qquad \times B_{\delta\delta\delta}(k_1,k_2,k_3,z) \; , 
\label{eq:defn_bispectrum}
\end{align}
where $\delta_{\rm D}$ is the Dirac-delta distribution. The statistical isotropy of the Universe implies that the power- and bispectrum only depend on the moduli of the $k$-vectors.

The projected power- and bispectrum can then be computed using the Limber approximation \citep{Limber:1954,Kaiser:1997,Bernardeau:1997,Schneider:1998,LoVerde2008},
\begin{equation}
P_{\kappa\kappa}^{(ij)} (\ell) = \int_0^{\chi_\mathrm{max}} \dd \chi\;\frac{g^{(i)}(\chi)\,g^{(j)}(\chi)}{a^2(\chi)}\, P_{\delta\delta}\left[\frac{\ell+1/2}{\chi},z(\chi)\right] \;, \label{eq:pkappa_defn}\\
\end{equation}
\begin{align}
B_{\kappa\kappa\kappa}^{(ijk)}(\ell_1,\ell_2,\ell_3) = {}&{} \int_0^{\chi_\mathrm{max}}\dd \chi\; \frac{g^{(i)}(\chi)\,g^{(j)}(\chi)\,g^{(k)}(\chi)}{a^3(\chi)\,\chi} \nonumber \\ {}&{}\times B_{\delta\delta\delta}\left[\frac{\ell_1}{\chi},\frac{\ell_2}{\chi},\frac{\ell_3}{\chi},z(\chi)\right] \label{eq:bkappa_defn}\, ,
\end{align}
where $\ell_3 = |\pmb{\ell}_1+\pmb{\ell}_2|$, and $g^{(i)}(\chi)$ denotes the lensing efficiency and is defined as
\begin{equation}
g^{(i)}(\chi) = \frac{3\Omega_\mathrm{m}H_0^2}{2c^2}\int_\chi^{\chi_\mathrm{max}} \dd \chi'\; n^{(i)}(\chi')\,\frac{\chi'-\chi}{\chi'} 
\label{eq:lensing_efficiency_defn} \, ,
\end{equation}
with $n^{(i)}(\chi)$ being the redshift probability distribution of the $i$-th tomographic $z_\mathrm{ph}$-bin.
Since the Limber approximation breaks down for small values of $\ell$ \citep{Kilbinger2017}, we consider only scales below $4^\circ$ to model $\MapMapMap$. In principle, we could have also used the modified Limber approximation and shift $\ell_{1,2,3}\rightarrow\ell_{1,2,3}+1/2$, but since we found a difference at maximum for the largest filter radii of $1.5\%$, we neglected it here. To model the non-linear matter power spectrum $P_{\delta\delta}$, we use the revised \textsc{HMcode2020} model of \citet{Mead2021} and for the matter bispectrum $B_{\delta\delta\delta}$ the \textsc{BiHalofit} of \citet{Takahashi:2020}.

\subsection{Non-linear alignment model}
\label{sec:IA_modelling}
The impact of galaxy IA is a known contaminating signal to the cosmic shear measurements that must be accounted for in all weak lensing studies. To model the effects of intrinsic alignment, we use the non-linear alignment (NLA) model \citep{Bridle2007}, which is a one-parameter model described as 
\begin{equation}
f_\mathrm{IA}=-\frac{A_{\mathrm{IA}}\bar{C_1}\bar{\rho}(z)}{D_+(z)} \, ,
\label{eq:_fIA}
\end{equation}
where $D_+$ is the linear growth factor at redshift $z$, $\bar{C_1}= 5\times 10^{-14} M_{\odot}^{-1} h^{-2}$  Mpc$^3$, as calibrated in \citet{Brown2002}, and $A_{\mathrm{IA}}$ captures the coupling strength between the matter density and the tidal field. 

By considering the tidal alignment field, $\delta_\mathrm{I}$, as a biased tracer of the matter density contrast field $\delta$, neglecting all higher order bias terms, we get:
\begin{equation}
\delta_\mathrm{I}(\boldsymbol{x},z) = f_\mathrm{IA} \delta(\boldsymbol{x},z) \, .
\end{equation}
With this, we find that:
\begin{align}
P_{\delta \delta_\mathrm{I}}(k,z)=f_\mathrm{IA}\, P_{\delta\delta}(k,z)\;, \\
P_{\delta_\mathrm{I} \delta_\mathrm{I}}(k,z)=f_\mathrm{IA}^2\, P_{\delta\delta}(k,z)\;,
\end{align}
where $P_{\delta\delta}(k,z)$ is the non-linear matter power spectrum.
The projected power spectra then follow to
\begin{align}
P^{(ij)}_\mathrm{GI}(\ell) &= \int_0^\infty \dd \chi \frac{g^{(i)}(\chi)\,n^{(j)}(\chi)}{a(\chi)\chi}\,P_{\delta \delta_\mathrm{I}}(\ell/\chi,\chi)\;, \\
P^{(ij)}_\mathrm{II}(\ell) &= \int_0^\infty \dd\chi \frac{n^{(i)}(\chi')\,n^{(j)}(\chi)}{\chi^2}\,P_{\delta_\mathrm{I} \delta_\mathrm{I}}(\ell/\chi,\chi)\;,
\end{align}
and the total projected power spectrum becomes 
\begin{equation}
P^{(ij)}(\ell) = P^{(ij)}_\mathrm{GG}(\ell) + P^{(ij)}_\mathrm{GI}(\ell) + P^{(ji)}_\mathrm{GI}(\ell) + P^{(ij)}_\mathrm{II}(\ell) \, .
\end{equation}
The term $P^{(ij)}_\mathrm{GG}(\ell)=P^{(ij)}_{\kappa \kappa}(\ell)$ describes the actual lensing signal, which is given in Eq.~\eqref{eq:pkappa_defn} with the weighting kernel in Eq.~\eqref{eq:lensing_efficiency_defn}. The II contribution describes how two galaxies spatially close together tend to be aligned. The term $\mathrm{GI}(z_i>z_j)$ describes the fact that high matter density regions align the lower redshift galaxies but also affect the shear of the background galaxies. While the II term is dominant if galaxies of the same tomographic bin are considered, GI and IG start to dominate if galaxies of separated tomographic bins are considered. With $z_i<z_j$, the term GI is expected to vanish for two tomographic bins with no significant redshift overlap.

Following the ansatz for modelling IA in the power spectrum, we get
\begin{align}
B_{\delta \delta \delta_\mathrm{I}}(k_1,k_2,k_3,z)&=f_\mathrm{IA}\, B_{\delta\delta\delta}(k_1,k_2,k_3,z) \, ,\\
B_{\delta \delta_\mathrm{I} \delta_\mathrm{I}}(k_1,k_2,k_3,z)&=f_\mathrm{IA}^2\, B_{\delta\delta\delta}(k_1,k_2,k_3,z)\, , \\
B_{\delta_\mathrm{I} \delta_\mathrm{I} \delta_\mathrm{I}}(k_1,k_2,k_3,z)&=f_\mathrm{IA}^3\, B_{\delta\delta\delta}(k_1,k_2,k_3,z)\, ,
\end{align}
where $B_{\delta\delta\delta}$ is the non-linear matter bispectrum, which we calculated with \textsc{BiHalofit}. The projected bispectra are
\begin{align}
&B^{(ijk)}_\mathrm{GGI}(\ell_1,\ell_2,\ell_3) = \nonumber \\ & \int_0^\infty \dd\chi \frac{g^{(i)}(\chi)\,g^{(j)}(\chi)\,n^{(k)}(\chi)}{a^2(\chi)\chi^2}\,B_{\delta \delta \delta_\mathrm{I}}(\ell_1/\chi,\ell_2/\chi,\ell_3/\chi,\chi)\, ,
\end{align}
\begin{align}
&B^{(ijk)}_\mathrm{GII}(\ell_1,\ell_2,\ell_3) = \nonumber\\ & \int_0^\infty \dd\chi \frac{g^{(i)}(\chi)\,n^{(j)}(\chi)\,n^{(k)}(\chi)}{a(\chi)\chi^3}\,B_{\delta \delta_\mathrm{I} \delta_\mathrm{I}}(\ell_1/\chi,\ell_2/\chi,\ell_3/\chi,\chi)\, ,
\end{align}
\begin{align}
&B^{(ijk)}_\mathrm{III}(\ell_1,\ell_2,\ell_3) = \nonumber \\ &\int_0^\infty \dd\chi \frac{n^{(i)}(\chi)\,n^{(j)}(\chi)\,n^{(k)}(\chi)}{\chi^4}\,B_{\delta_\mathrm{I} \delta_\mathrm{I} \delta_\mathrm{I}}(\ell_1/\chi,\ell_2/\chi,\ell_3/\chi,\chi) \, .
\end{align}
The total projected bispectrum is:
\begin{align}
B^{(ijk)}(\hat{\ell}) &=  B^{(ijk)}_\mathrm{GGG}(\hat{\ell}) + B^{(ijk)}_\mathrm{GGI}(\hat{\ell}) +
B^{(kij)}_\mathrm{GGI}(\hat{\ell}) +
B^{(jki)}_\mathrm{GGI}(\hat{\ell}) \nonumber \\ & +
B^{(ijk)}_\mathrm{GII}(\hat{\ell}) +
B^{(kij)}_\mathrm{GII}(\hat{\ell}) +
B^{(jki)}_\mathrm{GII}(\hat{\ell}) +
B^{(ijk)}_\mathrm{III}(\hat{\ell}) \, ,
\end{align}
where the tuple is $\hat{\ell}=(\ell_1,\ell_2,\ell_3)$. The interpretation of all these terms is analogous to the ones from the power spectrum, where $B^{(ijk)}_{GGG}(\hat{\ell})=B^{(ijk)}_{\kappa \kappa \kappa}(\hat{\ell})$ is the actual pure lensing signal given by Eq.~\eqref{eq:bkappa_defn} with the weighting kernel in Eq.~\eqref{eq:lensing_efficiency_defn}.

\subsection{Aperture mass statistics}
\label{subsec:background_map3}
One of the major problems of weak lensing mass reconstruction techniques is the mass-sheet degeneracy \citep[][hereafter MSD]{Falco:1985,1995A&A...294..411S}, which corresponds to adding a uniform surface mass density without affecting lensing observables such as the shear. However, it is possible to define quantities invariant under the MSD, one example being the aperture mass statistics \citep{Schneider:1996,Bartelmann:2001}. Another advantage of aperture mass statistics is that they separate the signal into so-called E- and B-modes \citep{Schneider:2002}, where, to leading order, the weak gravitational lensing effect cannot create B-modes. Lastly, as shown in 
\citetalias{Heydenreich2023}, aperture statistics are an excellent strategy to compress thousands of bins of 3PCF into a few hundred $\MapMapMap$ bins. 

The aperture mass $\Map$ at position $\pmb{\vartheta}$ with filter radius $\theta_\mathrm{ap}$ is defined through the convergence $\kappa$, as follows:
\begin{equation}
\label{eq:definition_aperture_masskappa}
\Map(\pmb{\vartheta};\theta_\mathrm{ap})=\int\dd^2\vartheta'\; U_{\theta_\mathrm{ap}}(|\pmb{\vartheta'}|)\, \kappa(\pmb{\vartheta}+\pmb{\vartheta'}) \; ,
\end{equation}
where $U_{\theta_\mathrm{ap}}(\vartheta')$ is a compensated filter such that ${\int \dd\vartheta'\,\vartheta'\, U_{\theta_\mathrm{ap}}(\vartheta') = 0}$. The tangential shear component $\gamma_\mathrm{t}$ of the complex shear in Cartesian coordinates $\gamma=\gamma_1 + \mathrm{i}\gamma_2$, is defined as $\gamma_\mathrm{t} = - \mathfrak{Re}(\gamma \mathrm{e}^{-2\mathrm{i}\phi})$, where $\phi$ is the polar angle of $\pmb{\vartheta'}$. Given the  tangential shear $\gamma_\mathrm{t}$, the aperture mass $\Map$ can also be calculated as 
\begin{equation}
\label{eq:definition_aperture_mass_gamma}
\Map(\pmb{\vartheta};\theta_\mathrm{ap})
=  \int\dd^2\vartheta' \; Q_{\theta_\mathrm{ap}}(|\pmb{\vartheta'}|)\,\gamma_\mathrm{t}(\pmb{\vartheta}+\pmb{\vartheta'}) \; ,
\end{equation}
where $Q_{\theta_\mathrm{ap}}$ is related to $U_{\theta_\mathrm{ap}}$ via
\begin{equation}
Q_{\theta_\mathrm{ap}}(\vartheta) = \frac{2}{\vartheta^2}\int_0^\vartheta \dd\vartheta'\;\vartheta'\, U_{\theta_\mathrm{ap}}(\vartheta')-U_{\theta_\mathrm{ap}}(\vartheta)\; .
\end{equation}
We define $U_{\theta_\mathrm{ap}}(\vartheta)=\theta_\mathrm{ap}^{-2}u(\vartheta/\theta_\mathrm{ap})$, denote by $\hat{u}(\eta)$ the Fourier transform of $u$ and use the filter function introduced in \citet{Crittenden:2002},
\begin{align}
u(x)= {}&{}\frac{1}{2\pi}\left(1-\frac{x^2}{2}\right)\ee^{-x^2/2}\;,\quad \hat{u}(\eta) = \frac{\eta^2}{2}\ee^{-\eta^2/2}\;, \nonumber\\
Q_{\theta_\mathrm{ap}}(\vartheta) = {}&{} \frac{\vartheta^2}{4\pi\theta_\mathrm{ap}^4}\exp\left(-\frac{\vartheta^2}{2\theta_\mathrm{ap}^2}\right)\; .
\end{align}

\subsection{Modelling aperture mass moments}
\label{subsec:modelling_map}

The expectation value of the aperture mass $\expval{\Map}(\theta_\mathrm{ap})$, which approximates the ensemble average over all positions $\pmb{\vartheta}$, vanishes by construction. However, the second-order (variance) of the aperture mass is nonzero and can be calculated as
\begin{equation}
\expval{\Map^2}^{(ij)}(\theta_\mathrm{ap}) = \int\frac{\dd \ell\;\ell}{2\pi}\,P^{(ij)}_{\kappa\kappa}(\ell)\,\hat{u}^2(\theta_\mathrm{ap}\ell)\; .
\end{equation}
Equivalently, the third-order moment of the aperture statistics $\MapMapMap$, can be computed from the convergence bispectrum via \citep{Jarvis:2004,Schneider:2005}
\begin{align}
&\expval{\Map^3}^{(ijk)}(\theta_{\mathrm{ap},1},\theta_{\mathrm{ap},2},\theta_{\mathrm{ap},3}) =  \int\frac{\dd^2 \ell_1}{(2\pi)^2}\int\frac{\dd^2 \ell_2}{(2\pi)^2}\ \nonumber \\ & \quad  \times B^{(ijk)}_{\kappa\kappa\kappa}(\ell_1,\ell_2,\ell_3)\,  \hat{u}(\theta_{\mathrm{ap},1}\ell_1) \, \hat{u}(\theta_{\mathrm{ap},2}\ell_2)\,\hat{u}(\theta_{\mathrm{ap},3}\ell_3)\; ,
\label{eq:Ma3_def}
\end{align}
where $\ell_3 = |\pmb{\ell}_1+\pmb{\ell}_2|$. Later, we differentiate between equal filter radii, for which $\theta_{\mathrm{ap},1}=\theta_{\mathrm{ap},2}=\theta_{\mathrm{ap},3}$ and non-equal filter radii, where the values of $\theta_{\mathrm{ap},i}$ are all allowed to vary.

\subsection{Modelling COSEBIs}
\citet{Schneider2010} introduced the complete orthogonal sets of E/B-integrals (COSEBIs), which are defined via the two-point shear statistics $\xi_{\pm}(\vartheta) = \expval{\gamma_\mathrm{t}\gamma_\mathrm{t}}(\vartheta) \pm \expval{\gamma_\mathrm{\times}\gamma_\mathrm{\times}}(\vartheta)$ on a finite angular range
\begin{align}
\label{eq:E_n_defintion}
E_n^{(ij)} &= \frac{1}{2} \int_{\vartheta_\mathrm{min}}^{\vartheta_\mathrm{max}} \dd \vartheta \,\vartheta \left[ T_{+n}(\vartheta) \xi_+^{(ij)}(\vartheta) + T_{-n}(\vartheta) \xi_-^{(ij)}(\vartheta) \right]\, , \\
\label{eq:B_n_defintion}
B_n^{(ij)} &= \frac{1}{2} \int_{\vartheta_\mathrm{min}}^{\vartheta_\mathrm{max}} \dd \vartheta \,\vartheta \left[ T_{+n}(\vartheta) \xi_+^{(ij)}(\vartheta) - T_{-n}(\vartheta) \xi_-^{(ij)}(\vartheta) \right]\, ,
\end{align}
where $T_{\pm n}(\vartheta)$ are filter functions with support in $[\vartheta_\mathrm{min},\vartheta_\mathrm{max}]$ \citep{Schneider2010}. If 
\begin{align}
\int_{\vartheta_\mathrm{min}}^{\vartheta_\mathrm{max}} \dd \vartheta \, \vartheta \, T_{+n}(\vartheta)\, J_0(\ell\,\vartheta) &= \int_{\vartheta_\mathrm{min}}^{\vartheta_\mathrm{max}} \dd \vartheta \, \vartheta \, T_{-n}(\vartheta)\, J_4(\ell\,\vartheta) \nonumber \\
&:= W_n(\ell) \, ,
\label{eq:W_n}
\end{align}
where $J_{0,4}$ are the zeroth and fourth order Bessel function, then the COSEBIs offer the advantage of cleanly separating all well-defined E- and B-modes within the range $[\vartheta_{\mathrm{min}},\vartheta_{\mathrm{max}}]$. This is not given, for instance, for second-order aperture mass statistics, as this would require information of $\xi_{\pm}$ over the full space. Analytically, the COSEBIs can be calculated from the E-mode power spectrum $P_{\kappa\kappa}^{(ij)} (\ell)$ defined in Eq.~\eqref{eq:pkappa_defn} and a B-mode angular power spectra $P_{\kappa\kappa,B}^{(ij)} (\ell)$ as:
\begin{align}
E_n^{(ij)} &= \int_{0}^{\infty} \frac{\dd \ell \, \ell}{2\pi} P_{\kappa\kappa}^{(ij)} (\ell) \, W_n(\ell) \, ,  \\
B_n^{(ij)} &= \int_{0}^{\infty} \frac{\dd \ell \, \ell}{2\pi} P_{\kappa\kappa,B}^{(ij)} (\ell) \, W_n(\ell) \, .
\end{align}
Since B-modes cannot be created by gravitational lensing directly, we neglected the modelling of $B_n$ for this work. Therefore, we focus, henceforth, only on the $E_n$ modes from the COSEBIs.

\section{Measuring shear statistics}
\label{subsec:measuring_map}

As discussed in \citetalias{Heydenreich2023} $\MapMapMap$ can be estimated from data in three ways, which we review below. The first uses the convergence field, $\kappa$, the second uses the shear field, and the third uses correlation functions. The latter is used to measure the real data and compute the covariance matrix used for the real data analysis.

\subsection{Measuring shear statistics from convergence maps}
\label{sec:Map_from^{(k)}appa}
If the convergence field, $\kappa$, is available (e.g. from simulations), the easiest way to measure $\Map$ is to use Eq.~\eqref{eq:definition_aperture_masskappa}. If the convergence field does not contain masks, this estimator is also unbiased as long as a border of the size of the filter function is removed from the aperture mass field before calculating the spatial average. However, no boundaries need to be removed for unmasked full-sky convergence fields to get an unbiased estimator. To estimate the aperture mass from (full-sky) convergence maps, the maps are smoothed with the \textsc{healpy}\footnote{\url{http://healpix.sourceforge.net}} \citep{healpy} function \textsc{smoothing}. This function needs a beam window function created by the function \textsc{beam$2$bl}, which is determined from the corresponding $U_{\theta_\mathrm{ap}}$ filter.

In the absence of B-modes the $E_n$ modes can also be calculated by using convolutions as \citep{Schneider2010}:
\begin{equation}
    E_n^{(ij)} = \langle \kappa^{(i)}(\pmb{\vartheta}) \kappa_s^{(j)}(\pmb{\vartheta};n) \rangle \; , 
\end{equation}
where 
\begin{equation}
\kappa_s^{(i)}(\pmb{\vartheta};n)=\frac{1}{2\pi}\int\dd^2\theta_\mathrm{ap}'\; T_{+n}(|\pmb{\vartheta}-\pmb{\vartheta'}|)\, \kappa^{(i)}(\pmb{\theta}_\mathrm{ap}') \,
\end{equation}
with $T_{+ n}$ introduced in Eq.~\eqref{eq:W_n}.

\subsection{Measuring shear statistics from galaxy catalogues}
\label{sec:Map_from_galcat_wo_mask}
In a real survey, the convergence is not observable and can be inferred only from the measured shear. However,  estimating $\Map$ from these reconstructed convergence maps is not accurate as the reconstructed convergence is necessarily smoothed and potentially also contains other systematic effects caused by masks or boundaries of the survey \citep{Seitz1997}. 

However, as motivated by Eq.~\eqref{eq:definition_aperture_mass_gamma}, the aperture mass can also be estimated from an unmasked shear field (e.g. for simulations). Given a galaxy catalogue, where galaxies are only found at specific positions such that the number of galaxies within an aperture varies on the sky, Eq.~\eqref{eq:definition_aperture_mass_gamma} needs to get modified to 
\begin{equation}
\label{eq:map_discrete_estimator}
    {\MapEst}^{(i)}(\boldsymbol{\vartheta};\theta_\mathrm{ap}) = \left[\sum_i Q_{\theta_\mathrm{ap}}(|\boldsymbol{\vartheta}-\boldsymbol{\vartheta}_i|)\right]^{-1} \sum_j Q_{\theta_\mathrm{ap}}(|\boldsymbol{\vartheta}-\boldsymbol{\vartheta}_j|)\,\varepsilon_{\mathrm{t},j}^{(i)}\; ,
\end{equation}
where the sum over the filter functions serves as a normalisation, $\varepsilon_{\mathrm{t}}$ are the observed galaxy ellipticities converted into their tangential component, and $\boldsymbol{\vartheta}_i$ are their respective positions. We note that we sampled the galaxies on a grid using a cloud-in-cell method\footnote{When placing the properties (shear) of a galaxy in a pixel grid, instead of shifting it to the pixel centre, we assume that the galaxy itself has the size of a pixel and distribute its properties to all neighbouring pixel centres weighted by their relative distance to the galaxy. This improves the accuracy on scales smaller than the pixel size.}, so that convolutions can determine the aperture masses. Since we used this approach solely for finite fields, we applied the same cut-off of $4\theta_\mathrm{ap}^\mathrm{max}$ to all aperture mass maps, where $\theta_\mathrm{ap}^\mathrm{max}$ is the largest filter radius. For both approaches, where the aperture mass is determined, the second- and third-order aperture statistics follow by multiplying the respective aperture mass maps pixel-wise and then taking the average of all pixel values.

\subsection{Measuring shear statistics from correlation functions} 
\label{sec:Map_from_3PCF}
The first two methods to estimate unbiased aperture statistics are not applicable to observed data with masks. Another disadvantage is the removal of the edges, which can be significant for a complex survey footprint, leading to decreased statistical certainty. The best method to estimate aperture shear statistics or the $E_n$ is by measuring them from two- and three-point shear correlation functions  \citep{Schneider:2002,Jarvis:2004,Schneider:2005}. The advantage of correlation functions is that they can be estimated for any survey geometry. The measurement of the two-point shear correlation functions $\xi_\pm (\theta_\mathrm{ap})$ is unbiased and easily performed by \textsc{treecorr} \citep{Jarvis:2004} and the conversion to COSEBIs are given in Eq.~\ref{eq:E_n_defintion} and Eq.~\ref{eq:B_n_defintion}. 

For the measurement of the aperture statistics, $\MapMapMap$, we refer to our companion paper, namely, \cite{Porth2023}. It describes an efficient estimation procedure of the natural components of the shear 3PCF \citep{Schneider:2003}, which is then transformed into aperture statistics. The corresponding equations for a tomographic set-up can be found in their Section 5.2.

\section{Observed data}
\label{Sect:real_Data}
This analysis explores the fourth data release of KiDS \citep{Kuijken:2015,Kuijken:2019,deJong:2015,deJong:2017}. The weak lensing data observed with the high-quality VST-OmegaCAM camera is public\footnote{The KiDS data products are public and available through \url{http://kids.strw.leidenuniv.nl/DR4}}. It is collectively called `KiDS-1000' as it covers $\sim 1000\,\mathrm{deg}^2$ of images, which is then reduced to an effective area of $777.4\,\mathrm{deg}^2$ after masking.
The significant advantage, compared to previous weak lensing surveys and data releases, is its overlap with its partner survey VIKING \citep[VISTA Kilo-degree Infrared Galaxy survey,][]{Viking2013}, which observes galaxy images at infrared wavelength. Therefore, galaxies were observed in nine optical and near-infrared bands, $u,g,r,i,Z,Y,J,H,K_\mathrm{s}$, allowing for better control over redshift uncertainties \cite[][hereafter \citetalias{Hildebrandt2021}]{Hildebrandt2021}.  

\begin{figure}[ht]
\includegraphics[width=\columnwidth]{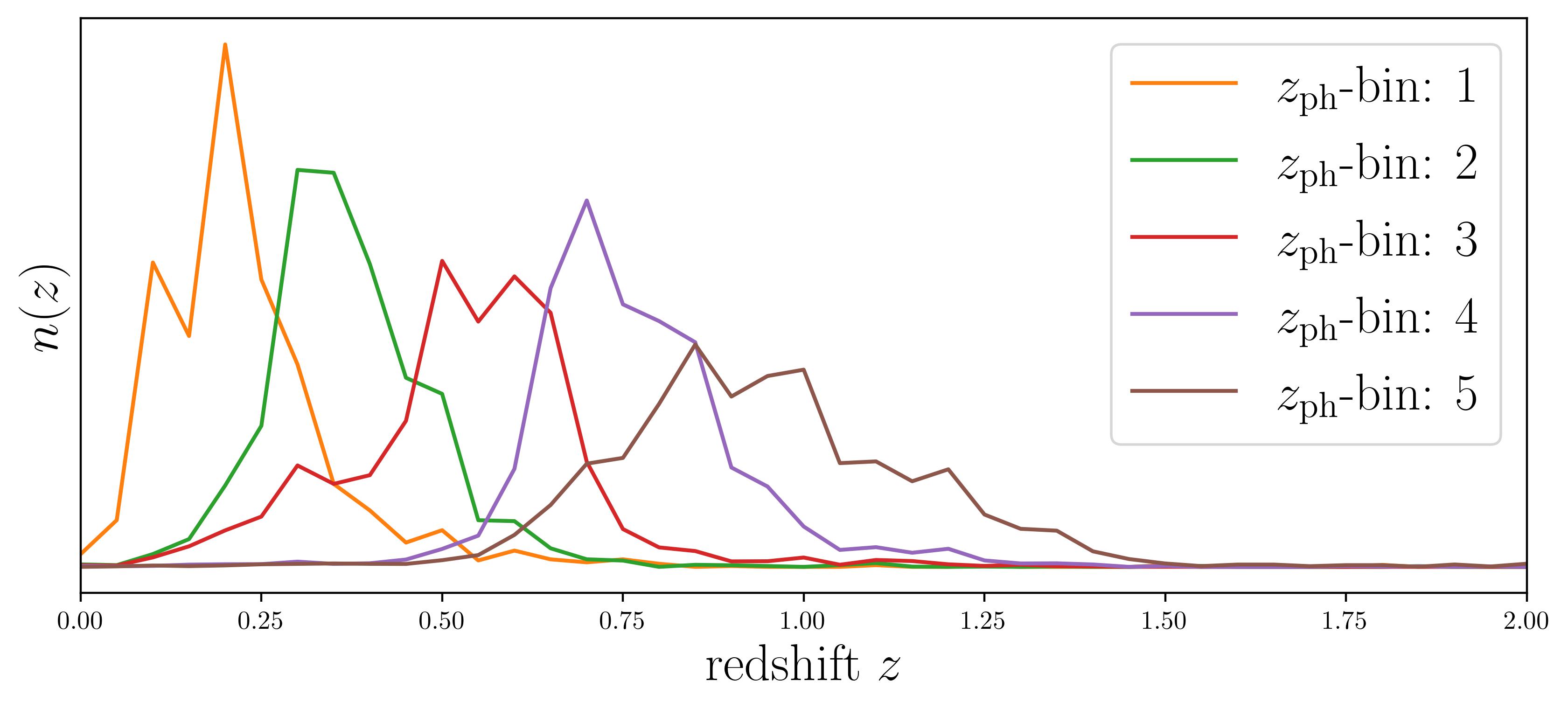}
\caption{Redshift distribution of the five tomographic $z_\mathrm{ph}$-bins of the KiDS-1000 data.}
\label{fig:nofz_real}
\end{figure}

The KiDS-1000 cosmic shear catalogue is divided into five tomographic $z_\mathrm{ph}$-bins, whose redshifts are calibrated using the self-organising map (SOM) method\footnote{By use of the nine-band photometry the SOM method allocates groups of galaxies to corresponding spectroscopic samples. If no matches are found, these galaxies are removed from the catalogue.} described in \citet{Wright:2020}. The redshift distributions of all five tomographic bins are shown in Fig.\,\ref{fig:nofz_real}, and were initially presented in \citetalias{Hildebrandt2021}. The residual systematic uncertainties on the redshift distributions are listed in Table \ref{table:data_overview} and are included in this work as nuisance parameters which we marginalise over. We note that they are correlated, which we account for by using their correlation matrix for the marginalisation. The galaxy shear ellipticities and their corresponding weights, $w$, are estimated by the lensfit tool \citep{Miller2013,Fenech2017,Kannawadi2019} and are described in more detail in \citet{Giblin:2020}. These shear-related systematic effects shift the $S_8$ parameter by (at most) $0.1\sigma$ when measured by cosmic shear two-point functions. The resulting systematics are stated in Table \ref{table:data_overview}, where we marginalise over the shear multiplicative $m$-bias correction in the resulting posteriors. 

\begin{table*}
\centering
\caption{Overview of the observational KiDS-1000 data.}
\begin{tabular}{ccccc}
\hline
\hline
name & $n_\mathrm{eff} [\mathrm{arcmin}^{-2}]$ & $\delta \langle z \rangle$ & $\sigma_\epsilon$ & $m$-bias $\times 10^{3}$\\
\hline
 $z_\mathrm{ph}$-bin: 1 & 0.62 & $0.000 \pm 0.0106$ &  0.270 & $-10\pm 19$  \\
$z_\mathrm{ph}$-bin: 2 & 1.18 & $0.002 \pm 0.0113$ &  0.258 & $-9\pm 20$  \\
$z_\mathrm{ph}$-bin: 3 & 1.85 & $0.013 \pm 0.0118$ &  0.273 & $-11\pm 17$  \\
$z_\mathrm{ph}$-bin: 4 & 1.26 & $0.011 \pm 0.0087$ &  0.254 & $8\pm 12$  \\
$z_\mathrm{ph}$-bin: 5 & 1.31 & $-0.006 \pm 0.0097$ & 0.270 & $12\pm 10$  \\
\hline
\hline
\vspace{0.1cm}
\end{tabular}
\tablefoot{The second column shows the effective number density, which also accounts for the lensfit weights. The third column shows the mean and uncertainty on the redshift bias taken from \citetalias{Hildebrandt2021}. The redshift bias uncertainties are correlated, which we accounted for by the correlation matrix given in \citetalias{Hildebrandt2021}. The fourth column displays the measured ellipticity dispersion per component, $\sigma_\epsilon$, as given in \citet{Giblin:2020}. The fifth column shows the shear multiplicative $m$-bias correction updated in \citet{JLB2022}.}
\label{table:data_overview}
\end{table*}

\section{Simulated data}
\label{Sect:Data}

We use several simulated data sets created to resemble the observed KiDS-1000 data, to validate our inference pipeline, study the impact of key systematic uncertainties, and forecast the expected KiDS-1000 analysis.\

In particular, we used the full-sky gravitational lensing simulations described in \citet[][hereafter  \citetalias{Takahashi2017}]{Takahashi2017} to generate data vectors and numerical covariance matrices. The cosmo-SLICS+IA simulations, described in \cite{Harnois-Deraps2022}, were used to test the modelling of IA; while the Magneticum lensing simulations, first introduced in \citet{Hirschmann2014}, were used to infuse Baryon feedback on the model whose strength we regulate with a free parameter in the posterior estimation.

\subsection{Takahashi simulations}
\label{sec:T17_description}

Since the \citetalias{Takahashi2017} simulations are used in this series of previous works, we only reiterate the essential details here.
The \citetalias{Takahashi2017} simulations follow the non-linear evolution of $2048^3$ particles evolved in a large series of nested cosmological volumes with side length starting at $L=450\,\mathrm{Mpc}/h$ at low redshift, and increasing at higher redshift, resulting in 108 different full-sky realisations. These were produced by the Gadget-3 $N$-body code \citep{Springel2005} and are publicly available\footnote{ \citetalias{Takahashi2017}  simulations: \url{http://cosmo.phys.hirosaki-u.ac.jp/takahasi/allsky_raytracing/}}. The cosmological parameters of the matter and vacuum energy density are fixed to $\Omega_{\rm m}=1-\Omega_\Lambda=0.279$, the baryon density parameter to $\Omega_{\rm b}=0.046$, the dimensionless Hubble constant to $h=0.7$, the normalisation of the power spectrum to $\sigma_8=0.82$, and the spectral index to $n_{\rm s}=0.97$. The shear information of the \citetalias{Takahashi2017} simulations is given in terms of 108 $\gamma$ and $\kappa$ full-sky realisations, where each realisation is divided into 38 ascending redshift slices. To reproduce the KiDS-1000 data for a given tomographic $z_\mathrm{ph}$-bin shown in Fig.~\ref{fig:nofz_real}, we built the weighted average of the first 30 $\gamma$ and $\kappa$ redshift slices. The weight for each redshift slice is measured by integrating the $n(z)$ over the corresponding width of the redshift slice.  

We decided on two approaches (convergence maps vs. galaxy ellipticity) to measure the data vectors and covariance matrices from the \citetalias{Takahashi2017} simulations. While the first approach (Sect.~\ref{sec:kappa_mock data}) has the advantage that a large number of mock data is available to measure a reliable covariance matrix even for large data vectors, the second approach (Sect.~\ref{sec:shear_mock data}) has the advantage that the exact galaxy positions are used, which implies that the holes (masks) match the data. 

\subsubsection{Convergence mock data}
\label{sec:kappa_mock data}
 The first approach is to convolve the full sky convergence maps with the $T_\mathrm{\pm,n}(\theta_\mathrm{ap})$ or $U(\theta_\mathrm{ap})$ filters, then multiply with corresponding other smoothed maps and then take the spatial average. For the covariance matrix, we divided the smoothed convergence maps into 48 sub-patches, where each patch has an area of $860\,\mathrm{deg}^2$. Since the $\kappa$ maps are first convolved with the filter functions and the sub-patches have common edges, the individual patches are not fully independent from each other, which decreases the covariance matrix. However, we found in our testing that selecting only 18 sub-patches, that have no common edges, gives almost identical results.
 Shape noise is added to the convergence maps by drawing random numbers from a Gaussian distribution with a vanishing mean and a standard deviation, as follows:
\begin{equation}
\label{eq:shapenoise}
    \sigma=\frac{\sigma_\epsilon}{\sqrt{n_\mathrm{eff}\, A_\mathrm{pix}}}\;,
\end{equation}
with the pixel area as $A_\mathrm{pix}=0.74\,\mathrm{arcmin}^2$ $(\textsc{nside}=4096)$, $n_\mathrm{eff}$ as the effective galaxy number density that included the lensfit weights, and $\sigma_\epsilon$ as the shape noise contribution given in Table \ref{table:data_overview}. With 48 sub-patches and 108 realisations, the covariance matrix is measured from 5184 mock data. As the actual KiDS-1000 area is roughly $777.4\,\mathrm{deg}^2$, we rescaled the covariance by $860\,\mathrm{deg}^2/777.4\,\mathrm{deg}^2\approx 1.1$. The reference data vector is measured from one \citetalias{Takahashi2017} realisation with a resolution of $A_\mathrm{pix}=0.18\,\mathrm{arcmin}^2$ $(\textsc{nside}=8192)$ without shape noise.

\subsubsection{Galaxy shear catalogue mock data}
\label{sec:shear_mock data}
For real surveys with a complex topology, the convolved maps give biased results \citep{Seitz1997}. Therefore, we decided on our second approach to measure our statistics from second- and third-order shear correlation functions. For this, we created galaxy shear catalogues from projected $\gamma$ and $\kappa$ fields by extracting the shear information only at the true positions of the observed galaxies in KiDS-1000. Since the correlation functions need to be measured to very small scales, we used realisations with a pixel resolution of $A_\mathrm{pix}=0.18\,\mathrm{arcmin}^2$ $(\textsc{nside}=8192)$. We checked that this resolution is sufficient by comparing the data vectors obtained from catalogues constructed from the same initial conditions on the reference resolution $(\textsc{nside}=8192)$ and a higher resolution $(\textsc{nside}=16384)$, finding a maximum deviation of $1\%$ for an aperture filter scale of $4'$. To increase the number of mock data, we shifted the galaxy positions 18 times by $\ang{20}$ along the lines of constant declination and extracted the shear information at the new positions. Afterwards, the shifted galaxy positions and the corresponding lensing information are shifted back to the original footprint. The back shifting is done only for simplicity and has no physical reason. With this procedure, 1944 almost independent mock data were created, from which the covariance and the reference data vector were measured. To add shape noise, we combined the two-component reduced shear $g = \gamma/(1-\kappa)$ of each object with a shape noise contribution, $\epsilon^{\mathrm{s}}$, to create observed ellipticities ${\boldsymbol\epsilon}^{\mathrm{obs}}$ \citep{Seitz1997}, as follows:
\begin{equation}
\epsilon^{\mathrm{obs}} = \frac{{ \epsilon}^{\mathrm{s}}+{ g}}{1+{ \epsilon}^\mathrm{s}{g}^*} \, .
\label{eq:ebos}
\end{equation}
The quantities here are all complex numbers and the asterisk `$*$' indicates complex conjugation. To mock the $\epsilon^{\mathrm{s}}$, we use the observed ellipticities $\epsilon_\mathrm{KiDS}^\mathrm{obs}$ of the KiDS-1000 data and randomly rotate them to erase the underlying correlated shear signal. This procedure offers the advantage that the resulting distribution of $\epsilon^{\mathrm{s}}$ matches the distribution of $\epsilon_\mathrm{KiDS}^\mathrm{obs}$. The estimated mean and dispersion $\sigma_\epsilon$ are given in Table \ref{table:data_overview}. Furthermore, when computing our statistics via correlation functions, we need to consider the corresponding weight $w$ for each shape measurement, which ensures that we use the correct effective number density. Since the lensing weights and the intrinsic ellipticities of source galaxies are correlated, adding the rotated ellipticities to the shear signal preserves this correlation.

\subsubsection{Data vector measurement and modelling}
\begin{figure*}[ht]
\includegraphics[width=\textwidth]{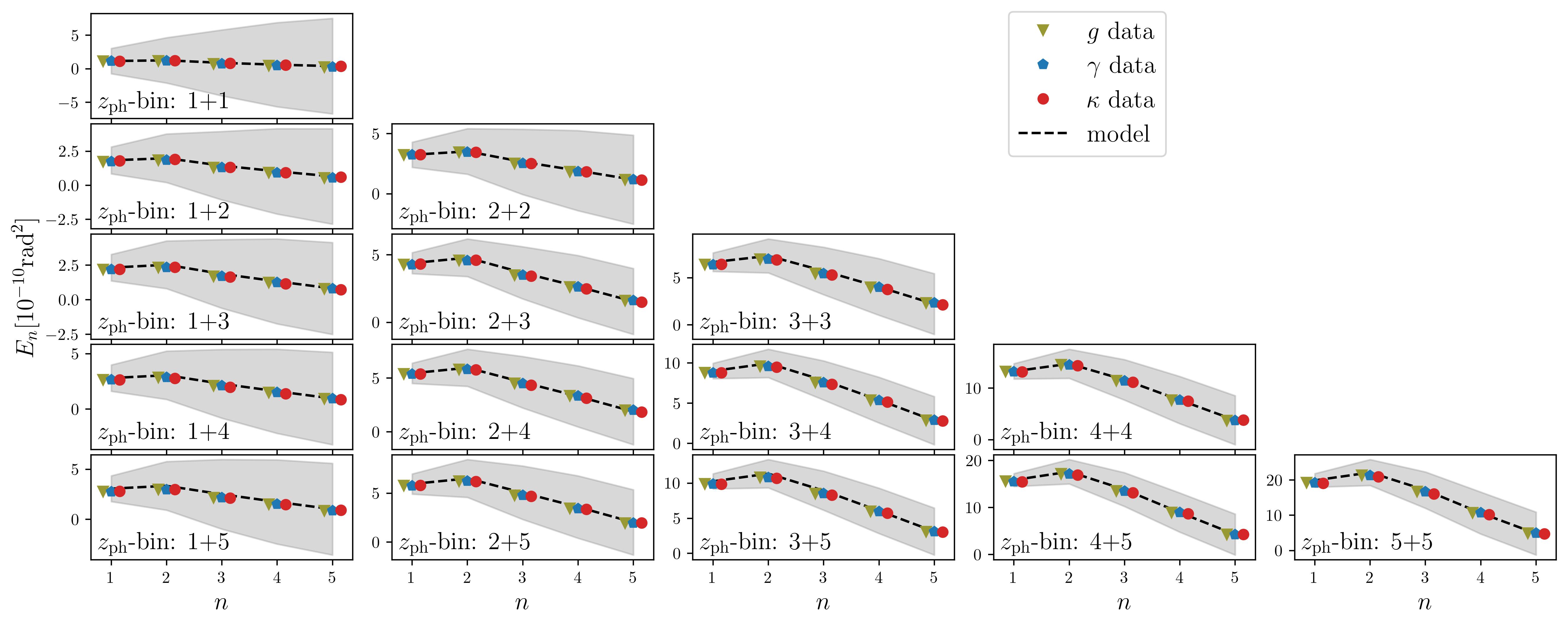}
\caption{Measured and modelled $E_n$ vector for the first five moments in the \citetalias{Takahashi2017} mock data. The green and blue dots are the mean of all 1944 KiDS-1000 mock data that are also used to compute the covariance matrix with a resolution $A_\mathrm{pix}=0.18\,\mathrm{arcmin}^2$ and shape noise. The red dots represent the data vector measured from one full-sky \citetalias{Takahashi2017} realisation with a resolution $A_\mathrm{pix}=0.18\,\mathrm{arcmin}^2$ and no shape noise. The grey band indicates the expected uncertainty from the KiDS-1000 survey.}
\label{fig:COSEBIs}
\end{figure*}

We follow the analysis of \citet[][hereafter \citetalias{Asgari2021}]{Asgari2021} as it is the fiducial cosmic shear analysis of the KiDS-1000\footnote{For updated KiDS-1000 cosmic shear analysis, we refer the reader to \cite{survey2023des} and \cite{Li2023}.}. We only use the first five $E_n$-moments determined from two-point correlation functions, measured from $\ang{;0.5}$ to $\ang{;300}$ in 400 radial bins. The resulting $E_n$ are shown in Fig.~\ref{fig:COSEBIs}, where we checked that the $B_n$ are consistent with zero. The $E_n$ data vector for $\gamma$\footnote{To compute the individual $\gamma$ data vectors with shape noise we have used Eq.~\eqref{eq:ebos} and replaced} and $g$ is the mean of all 1944 \citetalias{Takahashi2017} mock data; and for the $\kappa$ data vector, we used one realisation with resolution $A_\mathrm{pix}=0.18\,\mathrm{arcmin}^2$ $(\textsc{nside}=8192)$. The analytical model was computed using \textsc{CosmoSIS} \citep{Zuntz2015}.

\begin{figure*}[ht]
\includegraphics[width=\textwidth]{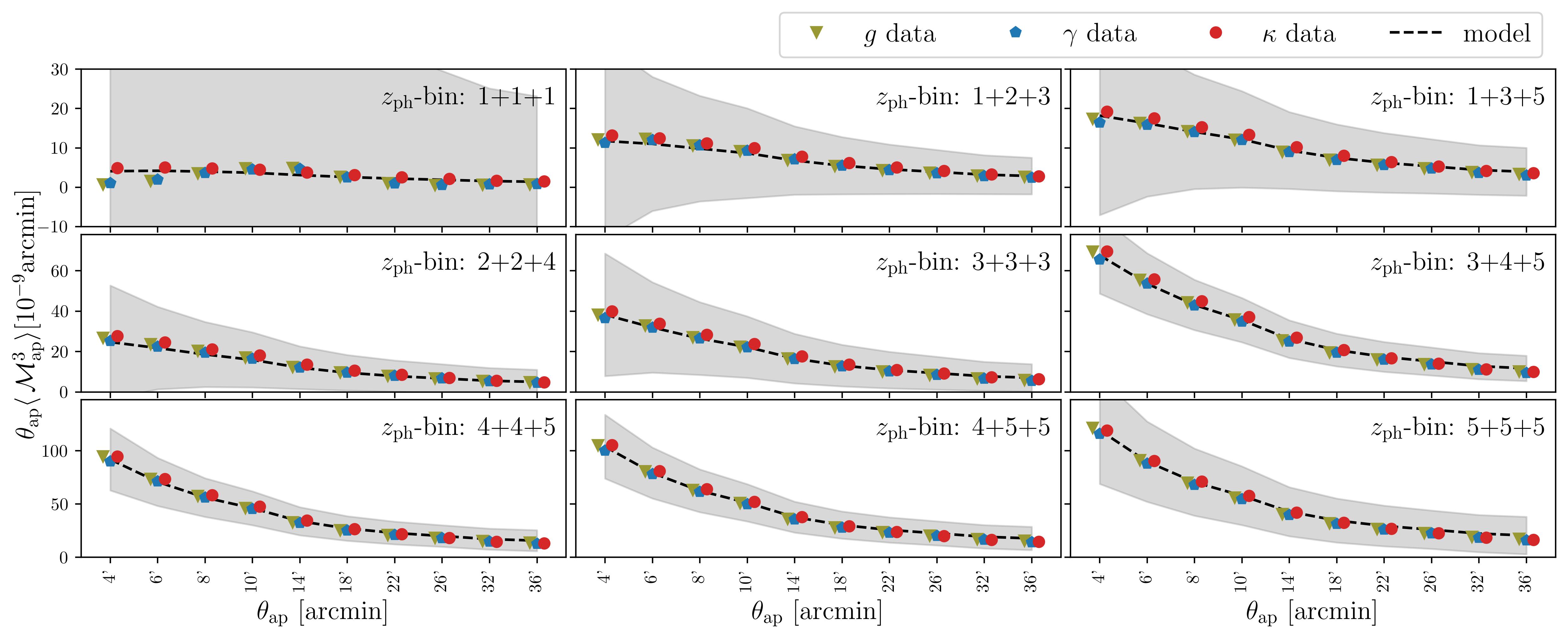}
\caption{Same as Fig.~\ref{fig:COSEBIs}, but here the measured data and model are the $\MapMapMap$ vector for equal-scale aperture filter radii $\theta_\mathrm{ap}\in\{\ang{;4}, \ang{;6}, \ang{;8}, \ang{;10},\ang{;14}, \ang{;18},\ang{;22}, \ang{;26}, \ang{;32}, \ang{;36}\}$ in the \citetalias{Takahashi2017} mock data for some selected $z_\mathrm{ph}$-bin combinations. The full data set is shown in Fig. \ref{fig:Map3_equalsclae}.
}
\label{fig:Map3_equalsclae_selected}
\end{figure*}

The $\MapMapMap$ data vectors are measured from the same mock data that are also used to compute the $E_n$. The analytical model pipeline is described in \citetalias{Heydenreich2023}. In Fig.~\ref{fig:Map3_equalsclae_selected}, we show $\MapMapMap$ for aperture filter radii $\theta_\mathrm{ap}\in\{\ang{;4}, \ang{;6}, \ang{;8}, \ang{;10},\ang{;14}, \ang{;18},\ang{;22}, \ang{;26}, \ang{;32}, \ang{;36}\}$, where we used only equal scales. We show them here for some specific $z_\mathrm{ph}$-bin combinations. In \ref{fig:Map3_allsclae}, we show $\MapMapMap$ also for non-equal aperture filter radii $\theta_\mathrm{ap}\in\{\ang{;4}, \ang{;8}, \ang{;16}, \ang{;32}\}$. 

\subsection{cosmo-SLICS+IA}
\label{sect:cosmo-SLICS+IA}

The cosmo-SLICS+IA simulations are cosmic shear mock galaxy catalogues infused with the non-linear alignment model of \citet{Bridle2007}, which is ideally suited for testing and validating our analytical IA model. They are based on $N$-body simulations of identical box size and particle density as the SLICS \citep{Harnois-Deraps:2018}, which were already used in previous works of this series; we therefore only summarise the essential details. 

The cosmology corresponds to the fiducial cosmo-SLICS model presented in \citet{Harnois-Deraps2019}, with $\Omega_{\mathrm{m}}=0.2905$, $\Omega_{\Lambda}=0.7095$,
$\Omega_{\mathrm{b}}=0.0473$, $h=0.6898$, $\sigma_8=0.836$, $w_0=-1.0$ and $n_{\rm s}=0.969$. We use the full set of 50 simulated galaxy catalogues, each covering 100 deg$^2$ and reproducing the KiDS-1000 $n(z)$ (see Fig.~\ref{fig:nofz_real}) and galaxy number densities $n_\mathrm{eff}$ specified in \citetalias{Hildebrandt2021}. As we use these mock data only to infuse IA effects into the \citetalias{Takahashi2017} data vector, shape noise is not included. Following the methods described in \cite{Harnois-Deraps2022}, the intrinsic ellipticity components $\epsilon_{1/2}^{\rm IA}$ of these galaxies are computed as
\begin{eqnarray}
\epsilon_1^{\mathrm{IA}} = - f_\mathrm{IA} (\partial_{xx}  - \partial_{yy})\phi \;,\quad    \epsilon_2^{\mathrm{IA}} = - 2 f_\mathrm{IA} \partial_{xy} \phi\, ,
\label{eq:tidal_th}
\end{eqnarray}
where $f_\mathrm{IA}$ is defined in Eq.~\eqref{eq:_fIA}, and $\phi$ is the gravitational potential. The partial derivatives of the gravitational potential describe the Cartesian components of the projected tidal field tensors. The $\epsilon_{1,2}^{\rm IA}$ terms were then combined with the noise-free cosmic shear signal using Eq.~\eqref{eq:ebos}, resulting in an IA-contaminated weak lensing sample that is consistent with the NLA model. As these mock data cover a square patch without any masking, we used the methods described in Sect.~\ref{sec:Map_from_galcat_wo_mask} to estimate the aperture statistics. Furthermore, they were used only to validate the IA modelling, which we discuss in Appendix \ref{sec:consistency_checks}.

\subsection{Magneticum simulations}
\label{sec:magneticum}
The feedback processes due to baryonic matter significantly affect the distribution of the LSS, such that the clustering of the matter is reduced on intra-cluster scales by up to tens of per cent \citep{vanDaalen2011}. However, in a quantitative sense, this suppression is not well understood \citep{Chisari2015}. We use the Magneticum simulations \citep{Hirschmann2014} to investigate the impact of baryonic feedback processes. Magneticum was run using {\sc Gadget3} code which is a more efficient version of Gadget 2 \citep{Springel2005} that includes modern smoothed particle hydrodynamics \citep{Beck2016}. The dark matter particle mass is $6.9 \times 10^8 h^{-1}M_\odot$ and gas particle mass $1.4 \times 10^8 h^{-1}M_\odot$. The underlying matter fields were constructed from the {\it Magneticum Pathfinder} simulations\footnote{\url{www.magneticum.org}} in the {\it Run-2} with a comoving volume of side $352\,h^{-1}\,{\mathrm{Mpc}}$ and {\it Run-2b} with a comoving volume of side $640\,h^{-1} \,{\mathrm{Mpc}}$, and are described in \citet{Hirschmann2014} and \citet{Ragagnin2017}, respectively. The Magneticum simulations account for radiative cooling, star formation, supernovae and active galactic nuclei.

The Magneticum shear maps used in this work were first presented in \cite{Castro2018}. Although the underlying creation of these shear maps is similar to the production of the cosmo-SLICS, meaning that the mock data follow the same redshift distribution and number density as the KiDS-1000 data, there are three main differences. First, only ten instead of 50 pseudo-independent light cones are available. Second, the galaxies are placed at the exact galaxy positions as in the observed data rather than randomly. Third, as these mock data are flat sky simulations and cover only an area of $100\,\mathrm{deg}^2$, the KiDS-1000 footprint is divided into 18 regions. This results in 18 catalogues for each of the ten pseudo-independent light cones. Due to the non-trivial geometry of the masks, we used correlation functions to estimate the shear statistics. 

To incorporate baryonic feedback processes into our modelling pipeline using the Magneticum simulations, we calculated the data vector, $\vb{d}^\mathrm{DM}$, using dark matter only and the data vector, $\vb{d}^\mathrm{DM+BA}$, where dark matter and baryons are included. With them we modified the model vector, $\vb{m}$, to
\begin{equation}
    \vb{m}' = \left[1+A_\mathrm{ba}\left(\frac{\vb{d}^\mathrm{DM+BA}}{\vb{d}^\mathrm{DM}}-1\right)\right]\vb{m} \, ,
    \label{eq:A_ba}
\end{equation}
where we introduce the parameter $A_\mathrm{ba}$. A value of zero for $A_\mathrm{ba}$ means no baryonic feedback, and a value of one is exactly the strength of baryonic feedback as in the Magneticum. Furthermore, we interpolate between zero and one and extrapolate to five, which we chose arbitrarily. We note that for the joint analysis of second- and third-order statistics, we need at least extrapolation to $A_\mathrm{ba}=2$ (at $95\%$ confidence), which approximately agrees with the baryonic strength of other hydrodynamical simulations \citep{Martinet2021A}. The third-order analysis alone is bound by the prior $A_\mathrm{ba}=5$. Although extrapolating to these large values is probably incorrect, we introduced $A_\mathrm{ba}$ only to give the model some flexibility to account for baryonic feedback effects without having a direct physical meaning such as the ejected gas.

\section{Cosmological parameter inference methodology}
\label{sec:inference_description}
In the following sections, we determine multiple posterior distributions using Markov chain Monte Carlo (MCMC) samplings for different model ingredients.  

We varied the two cosmological parameters $\Omega_\mathrm{m}$ and $S_8$ while fixing the remaining parameters to the  \citetalias{Takahashi2017} cosmology. Additionally, we varied the intrinsic alignment amplitude, $A_\mathrm{IA}$, and the parameter $A_\mathrm{ba}$ that accounts for the strengths of the response to baryonic feedback\footnote{This parameter is used for both the $E_n$ and the $\MapMapMap$ part of the data vector. We found that $A_\mathrm{ba}=[0.0,1.0,3.0]$ roughly corressponds to $\log_{10} T_\mathrm{AGN}=[7.1,7.6,8.5]$, which is the baryon parameter used in \textsc{HMcode2020}.} and was described in Sect.~\ref{sec:magneticum}. Lastly, we always marginalise over the shifts of the redshift distributions $\delta \langle z \rangle$ given in \citetalias{Hildebrandt2021} and the $m$-bias \citep{Giblin:2020} of all five tomographic bins shown in Table \ref{table:data_overview}. For these nuisance parameters, we assume Gaussian priors and account for the fact that the uncertainties of shifts of the redshift distributions are correlated using its correlation matrix. We give an overview of the priors in Table \ref{table:parameter_overview}.

\begin{table}[h!]
\centering
\caption{All varied parameters and their flat prior knowledge.}
\begin{tabular}{ccc}
\hline
\hline
parameter & validation & real \\
\hline
$\Omega_\mathrm{m}$ & $[0.10,0.50]$ & $[0.10,0.75]$  \\
$S_8$ & $[0.6,1.0]$ & $[0.6,1.0]$  \\
$A_\mathrm{IA}$ & $[-1.8,1.8]$ & $[-1.5,1.5]$  \\
$A_\mathrm{ba}$ & $[0.0,5.0]$ & $[0.0,5.0]$  \\
\hline
\hline
\end{tabular}
\tablefoot{Uniformly distributed  priors on the parameters used in our cosmological inferences. The priors given in Table \ref{table:data_overview} on the multiplicative shear $m$-bias and photometric redshift errors $\delta \langle z \rangle$ are used both for the real data analysis and for the mock data analysis, where for the latter the expectation values are set to zero. The $\delta \langle z \rangle$ for the sources follow a joint normal distribution with covariance matrix $C_{\delta \langle z \rangle}$ shown in figure 6 of \citetalias{Hildebrandt2021}. The $m$-bias follows an uncorrelated normal distribution. We increased the upper prior range on $\Omega_{\rm m}$ to 0.75 for the real data analysis to have $E_n$ posteriors that are not bound by the prior. The $A_\mathrm{IA}$ of the real data analysis are narrower to improve the emulator accuracy. The upper bound of $A_\mathrm{ba}$ is chosen arbitrarily and does play a role only for $\MapMapMap$-only analysis.}
\label{table:parameter_overview}
\end{table}

If the covariance matrix, $\Tilde{C}$, is measured from simulations, it is a random variable. We followed the method from \cite{Percival2021}, which leads to credible intervals that can also be interpreted as confidence intervals with approximately the same coverage probability. The posterior distribution of a model vector, $\vb{m}$, that depends on $n_{p}$ parameters, $\boldsymbol{p}$, if the covariance matrix $\Tilde{C}$ is measured from $n_\mathrm{r}$ mock data, is given by:
\begin{equation}
    \boldsymbol{P}\left[\vb{m}(\boldsymbol{p})|\vb{d},\Tilde{C}\right] \propto |\Tilde{C}|^{-\frac{1}{2}} \left( 1 + \frac{\chi^2}{n_{\rm r}-1}\right)^{-m/2}\, ,
    \label{eq:t_distribution}
\end{equation}
where $\vb{d}$ is the measured data vector and 
\begin{equation}
\chi^2 =  \left[\vb{m}(\boldsymbol{p})-\vb{d}\right]^{\rm T} \Tilde{C}^{-1} \left[\vb{m}(\boldsymbol{p})-\vb{d}\right] \, .
\label{eq:chi2}
\end{equation}
The power-law index $m$ is 
\begin{equation}
    m = n_p+2+\frac{n_\mathrm{r}-1+B(n_\mathrm{d}-n_p)}{1+B(n_\mathrm{d}-n_p)} \;,
\end{equation}
with $n_{\rm d}$ being the number of data points and
\begin{equation}
    B = \frac{n_\mathrm{r}-n_\mathrm{d}-2}{(n_\mathrm{r}-n_\mathrm{d}-1)(n_\mathrm{r}-n_\mathrm{d}-4)} \, .
    \label{eq:B}
\end{equation}
To give a quantified value for the comparison of different modelling choices, we use the Figure of Merit (FoM), which we calculate as
\begin{equation}
    \mathrm{FoM} = \frac{1}{\sqrt{\det C }} \, ,
\end{equation}
where $C$ is the parameter covariance matrix of the $S_8$-$\Omm$ plane resulting from the MCMC process.

Depending on the used filter combinations, the modelling of $\MapMapMap$ takes around $10$ to $20\,\mathrm{min}$, which stands as an obstacle to directly running an MCMC using the model. To circumvent this issue, we use the emulation tool contained in \texttt{CosmoPower} \citep{COSMOPOWER2022}. We trained the emulator on 3000 model points for the analysis based on mock data in the parameter space $\{\Omega_{\rm m}, S_8, A_\mathrm{IA} \}$ and 5000 model points for the real data analysis in the parameter space $\{\Omega_{\rm m}, S_8, A_\mathrm{IA}, \delta \langle z \rangle\}$, which we distributed in a Latin hypercube for the given prior range, and fixed all other parameters to the ones used in the \citetalias{Takahashi2017} simulations. The $m$-bias and the $A_\mathrm{ba}$ do not need to be emulated as they follow a simple correction formalism that is applied during the MCMC sampling. The $\delta \langle z \rangle$ of each $z_\mathrm{ph}$-bin for the training and testing are distributed uncorrelated between $[-0.06,0.06]$. The accuracy of the emulator is tested by comparing the emulator prediction with the model at 500 independent points in the same parameter space. The fractional error of each vector element is smaller than $2\%$ for $E_n$ and smaller than $5\%$ for $\MapMapMap$, which is well within the accuracy of the \textsc{HMcode2020} or the \textsc{BiHalofit} model itself. We used a Metropolis–Hastings sampler for MCMC\footnote{The code we made use of can be find here: \url{https://github.com/justinalsing/affine}}, where we used 1000 walkers running each 20\,000 steps and cut the first 2000 steps away to ensure that the posteriors are not biased by the burn-in phase. 

\section{Reduction of data vector combinations and filter radii}
\label{sec:combination_validation}

\begin{figure}[ht]
\includegraphics[width=\columnwidth]{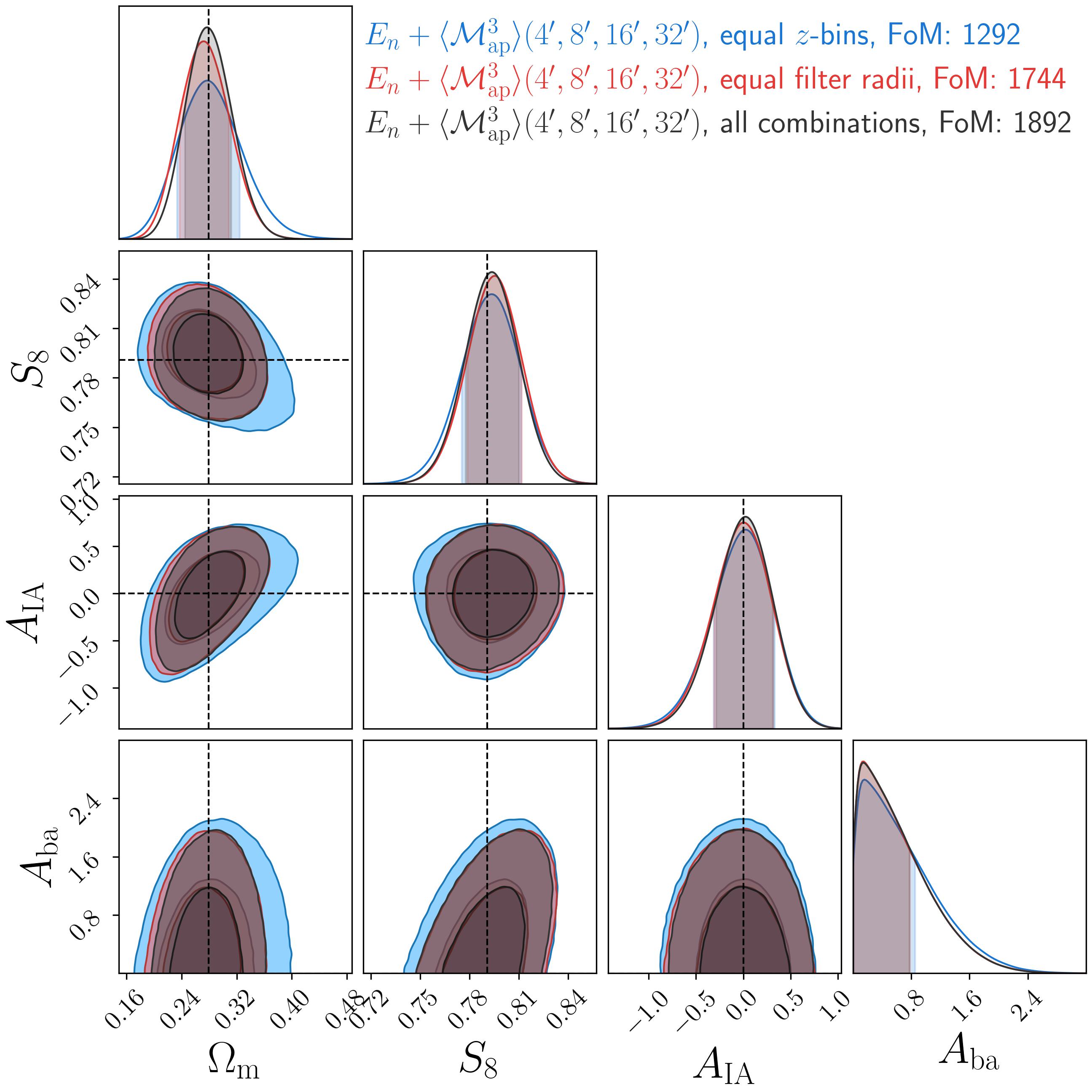}
\caption{Posterior distribution for data vectors and covariance measured from the \citetalias{Takahashi2017} convergence maps catalogues. Here only specific parts of the $\MapMapMap$ data vector are used. Here `only equal $z_\mathrm{ph}$-bins' means that only the auto tomographic bins are used. Lastly, all non-equal filter radii are discarded for `only equal filter radii'.}
\label{fig:MCMC_zbin_reduction}
\end{figure}

Generally, as long as the covariance is converged, the more information is used, the higher the constraining power. However, for third-order statistics with four different filter radii and five tomographic bins, the data vector contains 700 elements.
To obtain a reliable covariance matrix, roughly $\sim 10$ times the dimension of the data vector is needed. This implies that modelling and measuring the model/data vector and covariance matrix are time-consuming and require thousands of simulations. 
Therefore, it is inevitable for this and for future analyses to compress the data vector without substantial information loss. We note that this section is only concerned with reducing the dimension of the data vector and is not meant as a proper forecast, which we do in the following sections. The spatial resolution of the mock data used to compute the covariance induces an error smaller than $<1\%$ in the $E_n$ covariance matrix. However, as we only reduce the elements of $\MapMapMap$, while using all combinations of the $E_n$, this is unproblematic for the purpose of this section. We note again that the data vector was measured from one realisation with a resolution of $A_\mathrm{pix}=0.18\,\mathrm{arcmin}^2$, and that we use the $\kappa$ based covariance for this section to ensure the covariance matrix is converged even for the data vector with the largest dimension. As a first check, we show in Fig.~\ref{fig:MCMC_zbin_reduction} multiple element choices that discard a significant part of the $\MapMapMap$ data vector. Using only the five auto-tomographic bins is not a good choice for reduction, as seen in the fact that the FoM is reduced by $32\%$. However, using only equal-scale aperture radii is a better reduction, reducing the data vector to $28\%$ while reducing the FoM by only $\approx 8\%$. Using only equal-scale aperture radii also has the advantage of being modelled faster. Therefore, we continue using only the equal-scale aperture radii for the rest of this work, which reduces the dimension of the data vector from 775 to 215.

\begin{figure}[ht]
\includegraphics[width=\columnwidth]{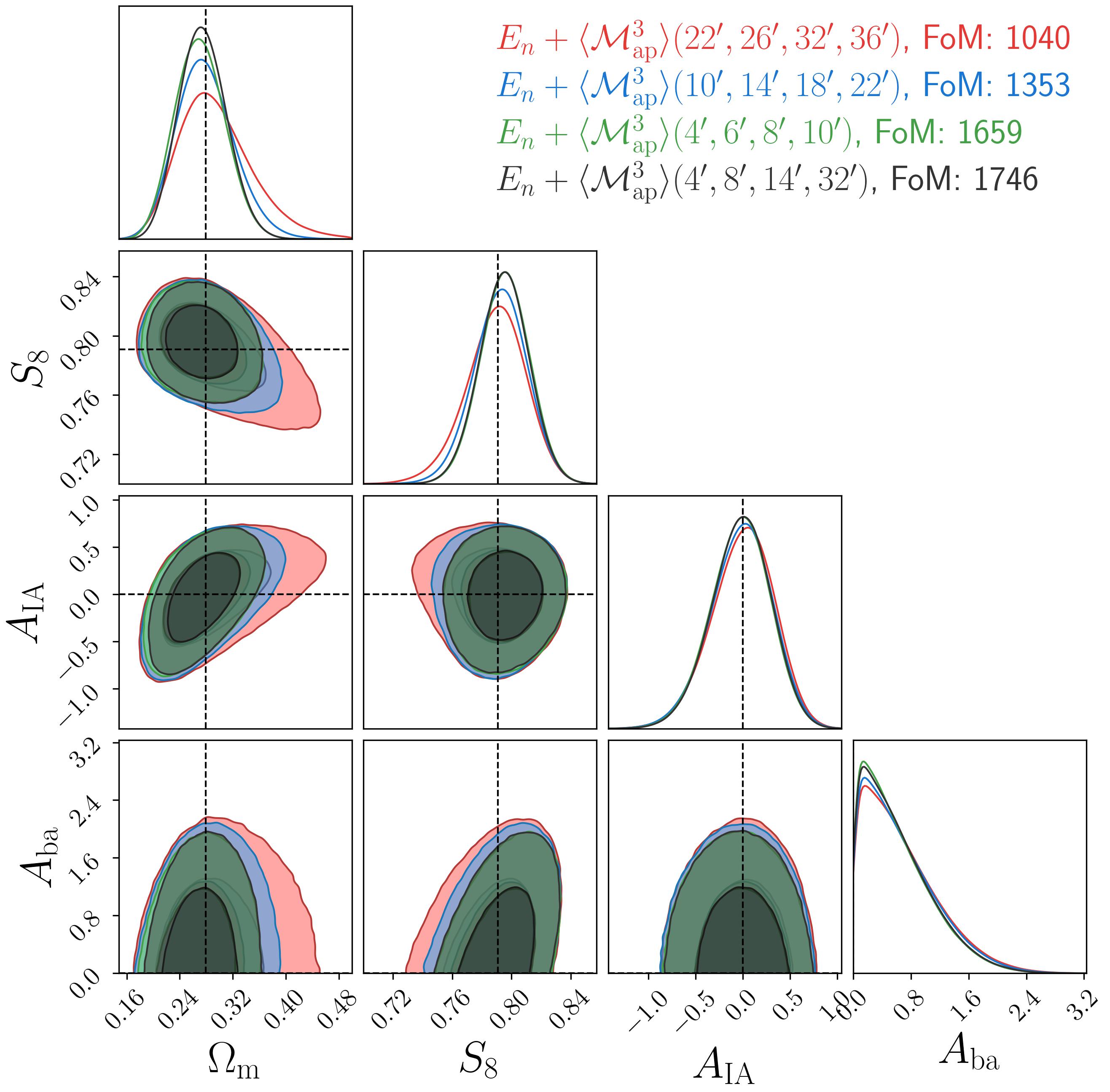}
\caption{Posterior distribution for different filter radii combinations if the data vector and covariances are measured from the \citetalias{Takahashi2017} convergence maps.}
\label{fig:MCCMC_combinations}
\end{figure}

Next, we investigated the choice of aperture filter radii. The aperture filter radii under consideration are $\theta_\mathrm{ap}\in\{\ang{;4}, \ang{;6}, \ang{;8}, \ang{;10}, \ang{;14}, \ang{;18}, \ang{;22}, \ang{;26}, \ang{;32}, \ang{;36}\}$. The resulting posteriors are displayed in Fig.~\ref{fig:MCCMC_combinations}, which reveals that using only the large filter radii above $>\ang{;22}$ have the worst constraining power. However, using only the four smallest filter radii is also not the best choice, as they are largely correlated and miss the large-scale information. The best choice of filter function is to use a filter from each range such as $\theta_\mathrm{ap}\in\{\ang{;4}, \ang{;8}, \ang{;14}, \ang{;32}\}$.

\begin{table}[h!]
\centering
\caption{Overview of the $\MapMapMap$ data vector choices and their relative constraining power.}
\begin{tabular}{c|c|c|c}
\hline
\hline
choice & Figure & rel. FoM & $n_d$ \\
\hline
all combinations & \ref{fig:MCMC_zbin_reduction} & 1.00  & 775 \\
only equal $z_\mathrm{ph}$-bins & \ref{fig:MCMC_zbin_reduction}  & 0.68 & 175 \\
$\theta_\mathrm{ap}\in\{\ang{;4}, \ang{;8}, \ang{;16}, \ang{;32}\}$  & \ref{fig:MCMC_zbin_reduction} & 0.92 & 215  \\
$\theta_\mathrm{ap}\in\{\ang{;4}, \ang{;8}, \ang{;14}, \ang{;32}\}$  & \ref{fig:MCCMC_combinations} & 0.92 &  215\\
$\theta_\mathrm{ap}\in\{\ang{;4}, \ang{;14}\}$  & - & 0.87 &  145\\
$\theta_\mathrm{ap}\in\{\ang{;4}, \ang{;14}, \ang{;32}\}$  & - & 0.89 &  180\\
$\theta_\mathrm{ap}\in\{\ang{;4}, \ang{;6}, \ang{;8}, \ang{;10}\}$  & \ref{fig:MCCMC_combinations} & 0.88  & 215\\
$\theta_\mathrm{ap}\in\{\ang{;10}, \ang{;14}, \ang{;18}, \ang{;22}\}$  & \ref{fig:MCCMC_combinations}& 0.72 & 215\\
$\theta_\mathrm{ap}\in\{\ang{;22}, \ang{;26}, \ang{;32}, \ang{;36}\}$  & \ref{fig:MCCMC_combinations} & 0.55  &  215 \\
all equal-filter radii  & - & 0.92 & 425\\
best 100 elements  & - & 0.80 &  100 \\
best 215 elements  & - & 0.89 &  200 \\
\hline
\hline
\end{tabular}
\tablefoot{For the measurement of the relative FoM the individual FoM is divided by the FoM of the $E_n + \MapMapMap$ of the `all combinations' case. From the second row onwards, only equal-scale filter radii are used.}
\label{table:Map3_choices}
\end{table}

Next, we considered the full data vector, meaning the $\MapMapMap$, and the $E_n$ for the next compression strategy. The idea is to decrease the number of elements by considering only those with the highest constraining power on $S_8$. We start with the element with the highest $S/N$, then consecutively add those vector elements that maximise the $S_8$ Fisher information content. This Fisher information is calculated as \citep{Tegmark1997}
\begin{equation}
    F_{S_8} = \left(\frac{\partial \pmb{m}(S_8)}{\partial S_8}\right)^T C^{-1} \left(\frac{\partial \pmb{m}(S_8)}{\partial S_8}\right) \;,
\end{equation}
where the partial derivatives are computed with a five-point stencil beam \citep{Fornberg:1988},
\begin{align}
   \frac{\partial \pmb{m}(S_8)}{\partial S_8} \approx \frac{-\pmb{m}(S_8^{++})+8\pmb{m}(S_8^{+})-8\pmb{m}(S_8^{-})+\pmb{m}(S_8^{--})}{12\Delta S_8}  \;,
\end{align}
where $S_8^{\pm} = S_8^\mathrm{T17} \pm 0.02 \, S_8^\mathrm{T17}$ and $S_8^{\pm\pm} = S_8^\mathrm{T17} \pm 0.04 \, S_8^\mathrm{T17}$. For this analysis, we used only the equal-scale filter radii $\theta_\mathrm{ap}\in\{\ang{;4}, \ang{;6}, \ang{;8}, \ang{;10}, \ang{;14}, \ang{;18}, \ang{;22}, \ang{;26}, \ang{;32}, \ang{;36}\}$. The first $\sim 200$ elements are sufficient to get converged posteriors. It is also interesting to see which elements help increase the constraining power. Unsurprisingly, the first elements are all COSEBIs, but among the first 100, approximately half are $\MapMapMap$ elements. Furthermore, it is interesting to see that cross-tomographic bin elements are more likely to be selected by our method, which is expected because they have a higher $S/N$ than auto-tomographic bin elements. Nevertheless, we also observe that using equal-scale filter radii $\theta_\mathrm{ap}\in\{\ang{;4}, \ang{;8}, \ang{;14}, \ang{;32}\}$  results in a better FoM then using the best 215 elements. This is likely because we optimised only the $S_8$ parameter here while fixing all others. Therefore, we also checked for the equal scale filter case if a principal component analysis (PCA) applied to the covariance matrix performs better in compressing the data. However, it needs more elements to get the same constraining power as our FoM maximiser. This is probably because a PCA considers only the covariance matrix and ignores the derivatives, meaning that a PCA is not necessarily sensitive to cosmology.

Finally, we give in Table \ref{table:Map3_choices} an overview of the FoM for the $\Omega_\mathrm{m}$--$S_8$ plane and the size of the data vector for some more element choices. As the covariance matrix is measured from 5184 mock data, we can assume that for all data vector dimensions under consideration, the covariance matrix is converged. Nevertheless, for the next sections, we use the covariance matrix measured from 1944 galaxy shear catalogues, limiting us to a maximum dimension of our data vector $\sim 200$ elements. Given the investigations in this section, we restrict the further sections to the case where we use all $E_n$ elements and all $\MapMapMap$ elements with the equal-scale filter radii $\theta_\mathrm{ap}\in\{\ang{;4}, \ang{;8}, \ang{;14}, \ang{;32}\}$, as this resulted in the best FoM for the restricted dimension of the data vector.

\begin{figure}[ht]
\includegraphics[width=\columnwidth]{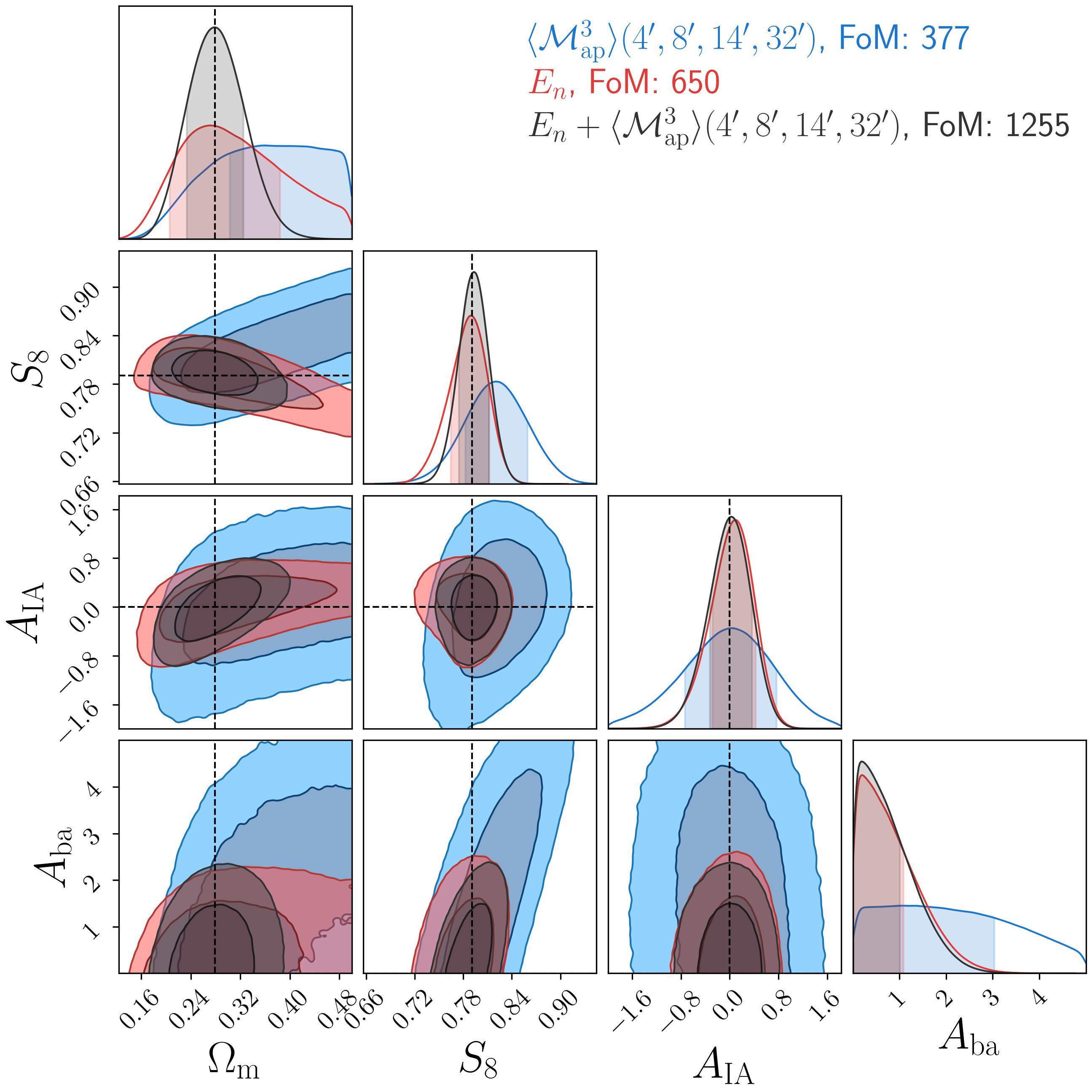}
\caption{ Posterior distribution for data vectors and covariance measured from the \citetalias{Takahashi2017} galaxy shear catalogues. The covariance matrix and reference data vector were measured from 1944 noisy $g$ mock data.}
\label{fig:MCCMC_estimator}
\end{figure}

\section{Validation of data vector estimator}
\label{sec:estimator_validation}
In a real survey analysis, two further difficulties arise. The first issue is that the lensing information is not given in terms of convergence maps but by point estimates (galaxies). The finite area where galaxies are measured implies that no lensing information is available outside that area. It is, therefore, necessary to measure the statistic from correlation functions described in Sect.~\ref{sec:Map_from_3PCF}. Second, these point estimates are the reduced shear $g=\gamma/(1-\kappa)$, which increases the signal and therefore needs to be accounted for. To correct these effects, we measured data vectors without shape noise and with the largest available resolution $A_\mathrm{pix}=0.05\,\mathrm{arcmin}^2$. Since these data vectors result only from one realisation without shape noise, we can expect that the ratio is a good approximation for the reduced shear effect. Although the deviations are small (as seen in Figs.~\ref{fig:Cosebis_Emodes_modelchecks} and \ref{fig:Map3_equalscale_modelchecks}), we decided to scale the model vectors by the ratio of $g$ and $\gamma$ data vector.

The resulting posteriors are shown in Fig.~\ref{fig:MCCMC_estimator}. Our first observation is that similar to the finding of \citetalias{Heydenreich2023}, combining second-order with third-order shear statistics significantly improves the constraining power on $S_8$ and $\Omega_\mathrm{m}$. Compared to the $E_n$-only, the $S_8$-$\Omega_\mathrm{m}$ FoM increases by $93\%$ and for the $\MapMapMap$-only by $233\%$. For these improvements, we have not considered that the posteriors of the individual statistics are bound by the priors on $\Omm$, meaning that the improvements are lower bounds. Compared to the analysis based on $\kappa$ mock data (see Fig.~\ref{fig:MCCMC_combinations}), the posteriors are broader because the covariance based on $\kappa$ maps also uses information outside the patches since the boundaries were not removed. This decreases the variance between the patches. A further difference is that the $\kappa$ analysis is not subject to masks and uses a slightly larger effective area. Although we rescaled the covariance to correct for the different effective areas, we found in \cite{Linke2023} that this rescaling is not necessarily accurate. Lastly, we notice that our modelling within the KiDS-1000 uncertainty is accurate, which we quantified by measuring the shift of the maximum of the posterior (MP) from the true values in the matter density parameter $\Delta \Omm < 0.02\, \sigma_{\Omm}$ and in the clustering amplitude $\Delta S_8 < 0.05\, \sigma_{S_8}$ with respect to the averaged noisy mock data vector and the KiDS-1000 uncertainty. We define the MP as the maximum of the one-dimensional marginal distributions.

\begin{figure*}[ht]
\includegraphics[width=\textwidth]{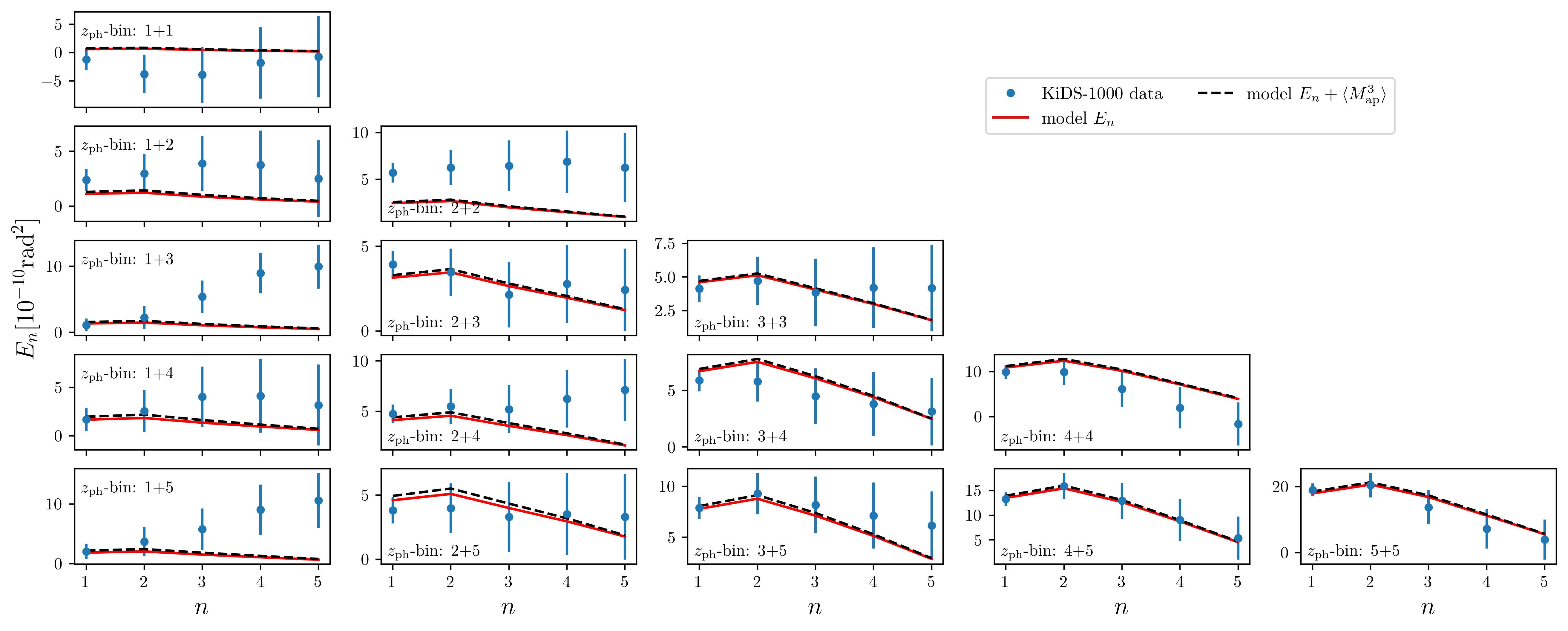}
\caption{Measured and modelled $E_n$ for the first five components. The blue points show the measurements from the KiDS-1000 data \citetalias{Asgari2021}. The blue error bars indicate the KiDS-1000 uncertainty. The different dashed lines show analytical descriptions at the MP if the combination of $E_n+\MapMapMap$ or only $E_n$ is used.}
\label{fig:Cosebis_Emodes_real}
\end{figure*}

\begin{figure*}[ht]
\includegraphics[width=\textwidth]{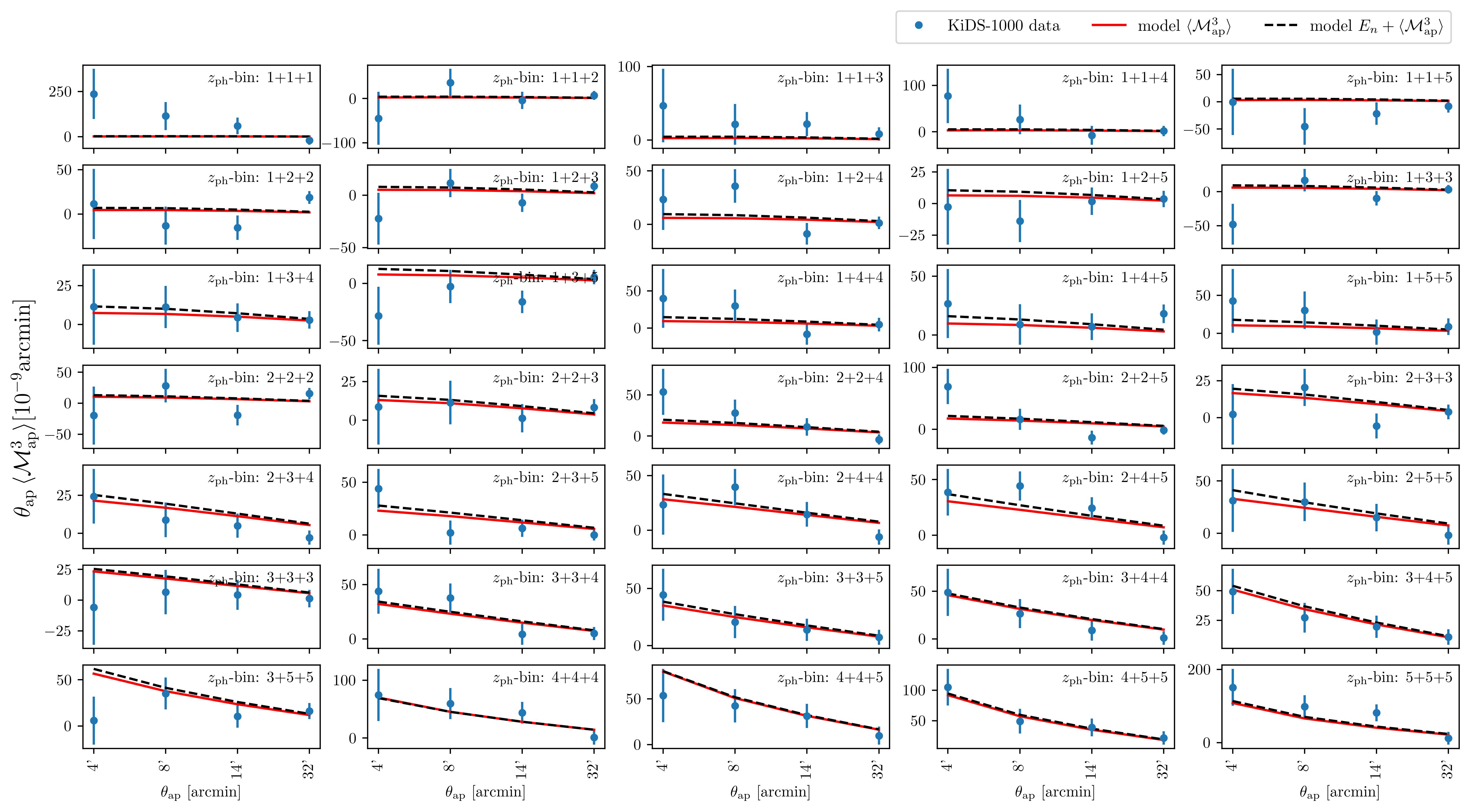}
\caption{Measured and modelled $\MapMapMap$ with filter radii $\theta_\mathrm{ap}\in\{\ang{;4}, \ang{;8}, \ang{;14}, \ang{;32}\}$. The blue error bars indicate the KiDS-1000 uncertainty. The different dashed lines show analytical descriptions at the MP if the combination of $E_n+\MapMapMap$ or only $\MapMapMap$ is used.}
\label{fig:Map3_equalscale_real}
\end{figure*}

\begin{figure*}[ht]
\includegraphics[width=\textwidth]{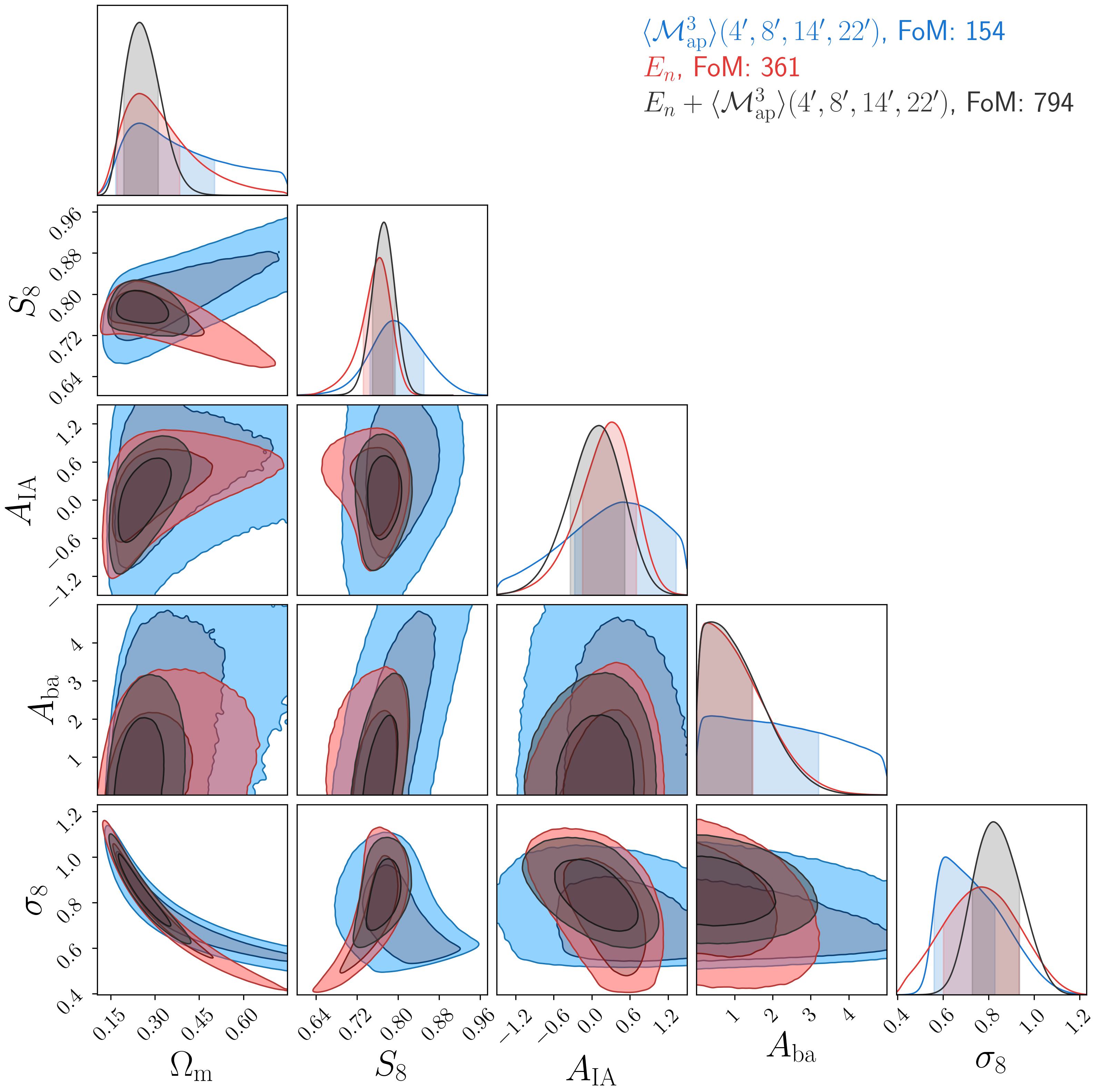}
\caption{Posterior distribution for the real KiDS-1000 data vector while the covariance is measured from the 1944 \citetalias{Takahashi2017} galaxy shear catalogues. Here the $E_n$ are compared to $\MapMapMap$ using all available combinations for the aperture filter radii $\theta_\mathrm{ap}\in\{\ang{;4}, \ang{;8}, \ang{;14}, \ang{;32}\}$ and redshift bin combinations. We note that the prior (see Table \ref{table:parameter_overview}) range on $\Omega_\mathrm{m}$ is enlarged compared to the previous validation plots. The FoM values of the $E_n$-only or $\MapMapMap$-only case should not be compared to Fig.~\ref{fig:MCCMC_estimator}, as priors of $\Omm$ bind their contours.}
\label{fig:MCMC_real}
\end{figure*}

\section{Cosmological results}
\label{sec:KiDS_results}

Finally, we are ready to present the first cosmological constraints from the KiDS-1000 data using second- and third-order shear statistics, displayed in Figs.~\ref{fig:Cosebis_Emodes_real} and \ref{fig:Map3_equalscale_real}. The data vectors are described in more detail in \citetalias{Asgari2021} for the $E_n$, and the $\MapMapMap$ in \cite{Porth2023}. Given the fact that the covariance matrix is measured only from 1944 realisations using the reduced shear $g$, we decide to build our data and model vector from all $E_n$ modes and $\MapMapMap$ with all equal filter radii $\theta_\mathrm{ap}\in\{\ang{;4}, \ang{;8}, \ang{;14}, \ang{;32}\}$. To control whether the model accurately fits the data, we minimised the $\chi_\mathrm{real}^2$ from the real data and the $\chi_\mathrm{T17}^2$ from each of the 1944 \citetalias{Takahashi2017} mock data used to compute the covariance matrix. To estimate the probability of measuring a $\chi^2>\chi_\mathrm{real}^2$ ($p$-value), we counted the number of $\chi_\mathrm{T17}^2$ that are greater than $\chi_\mathrm{real}^2$ divided by 1944. The resulting $p$-value for the $E_n$-only, the $\MapMapMap$-only, and the combination are given in Table \ref{table:MAP_values}. The resulting $p$-values for all combinations are better than 0.1, indicating that our covariance is well matched to the observed data and our model is accurate enough to describe the data. Our maximum posterior $\chi^2$ value increases to 81 if we swap our $E_n$ covariance matrix to the analytical expression in \cite{Joachimi2021} calculated at the MP parameter values in \citetalias{Asgari2021}. This is unsurprising as the numerical covariance is computed at the \citetalias{Takahashi2017} cosmology, giving a signal larger than that computed at the MP of \citetalias{Asgari2021}. Furthermore, our covariance matrix is measured from reduced shear mock data that slightly increases the covariance matrix. We also notice that the model and the data seem inconsistent for some $z_\mathrm{ph}$-bin combinations. However, adjacent COSEBI modes are highly correlated, so visually inspecting the model's goodness of fit to the data is misleading. We refer to \citetalias{Asgari2021} for further details on this discrepancy.

\begin{table*}[h!]
\centering
\caption{MP values with their marginal $68\%$ credible intervals.}
\begin{tabular}{c|ccccccc}
\hline
\hline
 & $\Omega_\mathrm{m}$ & $S_8$ & $\sigma_8$ &  $A_\mathrm{IA}$ & $A_\mathrm{ba}$ &  $\chi^2$ & $p$-value \\
\hline
$E_n$-only & $0.246^{+0.137}_{-0.076}$ & $0.765^{+0.025}_{-0.032}$ & $0.77^{+0.17}_{-0.16}$  & $0.31^{+0.39}_{-0.46}$ & $0.19^{+1.29}_{-0.18}$ & $72$ & $0.42$ \\
$\MapMapMap$-only & $0.248^{+0.253}_{-0.082}$ & $0.791^{+0.060}_{-0.046}$ & $0.61^{+0.22}_{-0.05}$ & $0.49^{+0.83}_{-0.76}$ & $0.39^{+2.83}_{-0.38}$ & $147$ & $0.25$ \\
$E_n + \MapMapMap$ & $0.248^{+0.062}_{-0.055}$ & $0.772\pm 0.022$ & $0.82^{+0.12}_{-0.09}$ & $0.12^{+0.40}_{-0.46}$ & $0.38^{+1.07}_{-0.37}$ & $224$ & $0.26$ \\
\hline
\hline
\end{tabular}
\tablefoot{The projected MP and their $68\%$ confidence intervals result from the MCMC chains shown in Fig.~\ref{fig:MCMC_real}. The values corresponding to the best $\chi^2$ might differ slightly. We fixed $h=0.6898$, $w_0=-1$ and $n_{\rm s}= 0.969$ but marginalised over the $\delta \langle z \rangle$ and $m$-bias uncertainties. The $p$-values were estimated by counting the \citetalias{Takahashi2017} mock data that give larger $\chi^2$ than the real $\chi^2$ relative to the total number of mock data.}
\label{table:MAP_values}
\end{table*}

We show the resulting posteriors in Fig.~\ref{fig:MCMC_real}, where we marginalised over the shift in the redshift distribution and multiplicative shear correction, both stated in Table \ref{table:data_overview}. We improve the constraints on $S_8$ by $23\%$ and on $\Omm$ by $47\%$ compared to the $E_n$-only case. This shows how powerful a combined analysis of the second- and third-order shear statistics is. The constraints on $A_\mathrm{IA}$ and $A_\mathrm{ba}$ are basically untouched, showing that $\MapMapMap$ is not helpful for constraining these nuisance parameters.

Compared to the maximum of the one-dimensional marginal distributions constraints of $E_n$ measurements given in \citetalias{Asgari2021} ($S_8 = 0.758^{+0.017}_{-0.026}$ and $\Omm=0.253^{+0.088}_{-0.074}$), we have slightly larger constraints when using only the $E_n$. This is because we use a numerical covariance, which is larger than the analytical one due to the underlying cosmology and the fact that we model the reduced shear effect. Furthermore, we note that \citetalias{Asgari2021} varied $\Omega_\mathrm{cdm}h^2$, $\Omega_\mathrm{b}h^2$ and $h$, while we fixed $h$ and $\Omega_\mathrm{b}$ and varied only $\Omm$. We also find that if we use the same pipeline as \citetalias{Asgari2021} but allow larger $\Omega_\mathrm{cdm}h^2$, the posteriors increase towards larger $\Omm$ and therefore get more consistent with our results. Furthermore, we use a different sampler compared to \citetalias{Asgari2021} and different baryon feedback process modelling. Nevertheless, our results from the $E_n$ analysis are consistent with \citetalias{Asgari2021} within $0.03\,\sigma$ in $\Omm$ and within $0.14\,\sigma$ in $S_8$. Similar to \citetalias{Asgari2021}, we also perform an internal consistency check, removing one $z_\mathrm{ph}$-bin at a time and finding consistent results. We discuss this in more detail in Appendix \ref{sec:consistency_checks}, and we find that (as expected) the fifth $z_\mathrm{ph}$-bin is most important to constrain $\Omm$ and $S_8$. We further discuss in Appendix \ref{sec:consistency_checks} modelling checks regarding the infusion of IA, baryon feedback and the reduced shear correction. The baryon feedback and reduced shear corrections are always sub-dominant compared to a shift in $S_8$. The IA, however, is important, especially if lower $z_\mathrm{ph}$-bins are included and must be accurately modelled.

Finally, we notice that all statistics are consistent with $A_\mathrm{IA}=0$, although both statistics alone seem to favour positive $A_\mathrm{IA}$. Interestingly, the joint analysis is shifted to lower $A_\mathrm{IA}$ and is slightly more constraining, which we do not observe in the validation in Sect.~\ref{sec:estimator_validation} where all posteriors accurately peak at the input value and all constraining power comes from the $E_n$. This might indicate that for real data with non-zero $A_\mathrm{IA}$, third-order shear statistics can contribute (at least a bit) to constraining IA, in line with predictions by, for instance, \cite{Pyne:2021} and \cite{Troxel2012}.  
Furthermore, the baryonic parameter can be confined to $A_\mathrm{ba}<1.4$ at $68\%$ confidence with our statistic. Here, we should note that $\MapMapMap$ does not help in constraining $A_\mathrm{ba}$, which is probably due to the fact that changes in $S_8$ absorb all $A_\mathrm{ba}$ effects.

\section{Conclusions}
\label{sec:Conclusions}
This work validates the combined modelling of second- and third-order shear statistics, showing that its accuracy is well-suited for a KiDS-1000 analysis. Our second-order shear statistic of choice is the $E_n$-modes of the COSEBIs \citep{Schneider2010} and the third-order statistic is $\MapMapMap$ \citep{Schneider:2005}. In particular, we incorporated intrinsic alignment modelling based on the non-linear alignment model of \cite{Bridle2007} and validated its accuracy against simulations infused with IA effects. This test is also interesting for other simulation-based analyses for which IA cannot be modelled analytically. We incorporated the impact of the baryonic feedback process by measuring a response function using the Magneticum simulations. Since the amplitude of this response function has no physical meaning, it is considered a nuisance parameter, which does not bias our cosmological parameter predictions.  

We investigate which parts of the data vector can be neglected without losing too much cosmological information. This is important because the data vector for third-order shear statistics, due to the possibility it offers of combining three different filters with three different redshift bins, inflating the data vector to several hundred elements easily. Therefore, both a numerical and an analytical covariance are difficult to compute. We find that cross-tomographic redshift bins contain a large amount of cosmological information. Using only equal-scale filter radii but all available tomographic bin combinations was the best data compression strategy, which comes with the advantage that equal-scale filters are faster to compute analytically as they require a lower integral accuracy. Next, we investigated the chosen filter radii. For this we measured and modelled the $\MapMapMap$ for $\theta_\mathrm{ap}\in\{\ang{;4}, \ang{;6}, \ang{;8}, \ang{;10},\ang{;14}, \ang{;18},\ang{;22}, \ang{;26}, \ang{;32}, \ang{;36}\}$. We find that filter radii above $\ang{;30}$ do not contribute much constraining power and that having many small filter radii is unnecessary. The best option is to use filter radii $\theta_\mathrm{ap}\in\{\ang{;4}, \ang{;8},\ang{;14}, \ang{;32}\}$. We also tested whether selecting elements that maximise the Fisher information matrix on $S_8$ is more optimal but find no difference compared to selecting all equal-scale filter radii. However, this method can be used to speed up the modelling and measurement of future analyses by discarding irrelevant elements. 

Next, we validated the assumption that the correlation function estimator gives accurate results. This is important since only correlation function estimators have the potential to give unbiased results for real data. We used realistic mock data created from the \citetalias{Takahashi2017} simulations. In particular, we created several galaxy catalogues where the positions of the galaxies are exactly at the KiDS-1000 galaxy positions. We had to rely on correlation functions to measure the second- and third-order statistics, which give unbiased results also if the data has a complex topology. Our first finding is that our modelling and measurement result in unbiased cosmological parameters given the KiDS-1000 uncertainty. Second, we find that using the reduced shear, or the shear itself, does not change the results and can, therefore, be ignored for the KiDS-1000 data analysis. 

We conclude this paper with an analysis of the real KiDS-1000 data. Overall, our $E_n$ constraints are less informative than the original KiDS-1000 analysis. This is mostly because we used a numerical covariance matrix from \citetalias{Takahashi2017} simulations. However, the chosen sampler, the priors on the cosmological parameters, and the modelling strategy of the baryonic feedback processes impact our constraints, too.
We find an $S_8=0.772\pm 0.022$ and an $\Omm=0.248^{+0.062}_{-0.055}$, which are improved compared to the $E_n$-only case by $23\%$ and by $47\%$, respectively. With a $p$-value of 0.25, we also find a good agreement of model and data given the KiDS-1000 uncertainty. This demonstrates that combining second- and third-order statistics is powerful in constraining cosmological parameters. The gain in constraining power in $\Omm$ is also interesting for combined weak lensing and galaxy clustering analysis because the constraining power in $\Omm$ for clustering analysis comes with the issue of further nuisance parameters such as galaxy bias. However, since second- and third-order shear statistics constrain $\Omm$ quite well, combining it with clustering statistics might enable us to learn more about these nuisance parameters. 

We leave the optimisation of the IA and baryonic feedback modelling for future analysis. An especially interesting improvement would be a more physically motivated description of the baryon feedback processes, which are identical for power and bispectrum. As all baryon feedback models rely on hydro simulations, we have to use the same simulations for power and bispectrum. Furthermore, although we found that the reduced shear and limber approximation is sufficient for a KiDS-1000 analysis, we probably have to model these effects for future Stage IV surveys. Lastly, we ignore the effect of source clustering for this work. Although \cite{Gatti2023_SC} found it relevant for third-order weak lensing statistics based on convergence mass maps, we expect it to be less critical for our analysis, which uses shear catalogues and no mass map reconstruction. For future Stage IV surveys, this needs to be investigated.

\begin{acknowledgements}
This paper went through the KiDS review process, and we want to thank the KiDS internal reviewer for the fruitful comments to improve this paper. We thank Mike Jarvis for maintaining \texttt{treecorr} and his constructive and fruitful comments. Furthermore, we thank Alessio Mancini for developing the \texttt{CosmoPower} emulator, which made the analysis pipeline of this work significantly faster.
 
Some of the results in this paper have been derived using the \textsc{healpy} and \textsc{healix}\footnote{currently http://healpix.sourceforge.net} package \citep{Gorskietal2005, healpy}. The figures in this work were created with {\sc matplotlib} \citep{hunter:2007} and {\sc ChainConsumer} \citep{hinton:2016}. We further make use of {\sc CosmoSIS} \citep{zuntz/etal:2015}, {\sc numpy} \citep{harris/etal:2020} and {\sc scipy} \citep{jones:2001} software packages.
LP acknowledges support from the DLR grant 50QE2002. SH is supported by the U.D Department of Energy, Office of Science, Office of High Energy Physics under Award Number DE-SC0019301. We acknowledge use of the lux supercomputer at UC Santa Cruz, funded by NSF MRI grant AST 1828315. LL is supported by the Austrian Science Fund (FWF) [ESP 357-N]. TC are supported by the INFN INDARK PD51 grant and by the FARE MIUR grant `ClustersXEuclid' R165SBKTMA. KD acknowledges support by the COMPLEX project from the European Research Council (ERC) under the European  Union’s Horizon 2020 research and innovation program grant agreement ERC-2019-AdG 882679 as well as support by the Deutsche Forschungsgemeinschaft (DFG, German Research Foundation) under Germany’s Excellence Strategy - EXC-2094 - 390783311. JHD acknowledges support from an STFC Ernest Rutherford Fellowship (project reference ST/S004858/1). HH is supported by a DFG Heisenberg grant (Hi 1495/5-1), the DFG Collaborative Research Center SFB1491, as well as an ERC Consolidator Grant (No. 770935). KK acknowledges support from the Royal Society and Imperial College. NM acknowledges support from the Centre National d’Etudes Spatiales (CNES) fellowship.\\

Author contributions: All authors contributed to the development and writing of this paper. The authorship list is given in three groups: the lead authors (PAB, LP, SH, LL, NW, PS), followed by an alphabetical group. 
This alphabetical group includes those who have either made a significant contribution to the data products or to the scientific analysis.\\

The KiDS results in this paper are based on observations made with ESO Telescopes at the La Silla Paranal Observatory under programme IDs 177.A- 3016, 177.A-3017, 177.A-3018 and 179.A-2004, and on data products produced by the KiDS consortium. The KiDS production team acknowledges support from: Deutsche Forschungsgemeinschaft, ERC, NOVA and NWO-M grants; Target; the University of Padova, and the University Federico II (Naples). Data processing for VIKING has been contributed by the VISTA Data Flow System at CASU, Cam- bridge and WFAU, Edinburgh.

The calculations for the hydrodynamical simulations were carried out at the Leibniz Supercomputer Center (LRZ) 
under the project pr83li (Magneticum). JHD acknowledges support from an STFC Ernest Rutherford Fellowship (project reference ST/S004858/1).

\end{acknowledgements}

\bibliography{bibliography.bib}

\begin{appendix}

\section{Consistency checks and modelling choices}
\label{sec:consistency_checks}
\begin{figure}[ht]
\includegraphics[width=\columnwidth]{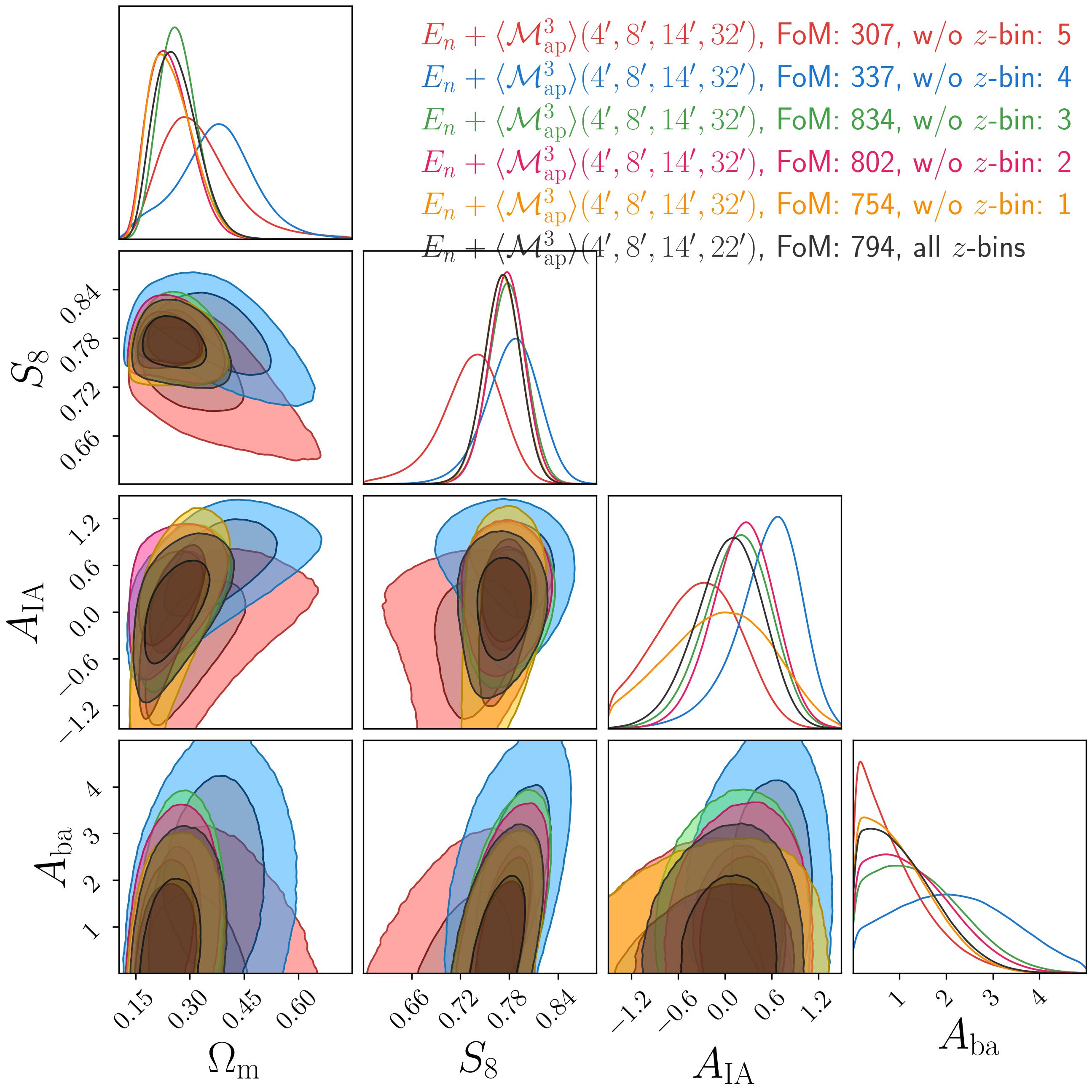}
\caption{Same as black contours in Fig.~\ref{fig:MCMC_real} but while removing one $z_\mathrm{ph}$-bin at a time.}
\label{fig:MCCMC_wozbins}
\end{figure}

As a first consistency check, we again inferred the joint analysis posteriors but removed one $z_\mathrm{ph}$-bin at a time. The results are shown in Fig.~\ref{fig:MCCMC_wozbins}
and reveal (most importantly) that all $z_\mathrm{ph}$-bins are internally consistent with each other. However, we further observe that $z_\mathrm{ph}$-bin one and $z_\mathrm{ph}$-bin two are not important for inferring cosmological parameters, as expected, given their S/N as shown in \cite{Porth2023}. For the $A_\mathrm{IA}$ parameter, in turn, $z_\mathrm{ph}$-bin one is very important, as the low redshift is most sensitive to IA. Next, we observe that the third $z_\mathrm{ph}$-bin results are basically the same as for all five $z_\mathrm{ph}$-bins, meaning that the third bin can eventually be discarded in future KiDS analyses if the dimension needs to be decreased further. Lastly, if either the fourth or fifth $z_\mathrm{ph}$-bin is removed, the constraining power on $\Omm$ and $S_8$ drastically decreases. For the fifth $z_\mathrm{ph}$-bin, this also affects the constraining power on $A_\mathrm{IA}$. 

Now we investigate the accuracy of the modelling strategy of IA described in Sect.~\ref{sec:IA_modelling}. To validate our IA modelling, we measured the data vector from the cosmo-SLICS+IA for $A_\mathrm{IA}=\{-1,0,1\}$ and then took the ratio of the data vector at $A_\mathrm{IA} \in \{-1,1\}$ and divided it by the data vector at $A_\mathrm{IA}=0$. This ratio is multiplied with the averaged \citetalias{Takahashi2017} $\gamma$ mock data vector to infuse IA. For this analysis, we used only equal-scale filter radii $\theta_\mathrm{ap}\in\{\ang{;4}, \ang{;8}, \ang{;14}, \ang{;32}\}$. As shown in Fig.~\ref{fig:MCMC^{(i)}A}, our model nicely recovers the input IA amplitude without shifting the other parameter posteriors. Interestingly, the larger $A_\mathrm{IA}$ the larger the FoM, which is due to the increased degeneracy breaking seen in the $\Omm$-$A_\mathrm{IA}$ panel.

\begin{figure}[ht]
\includegraphics[width=\columnwidth]{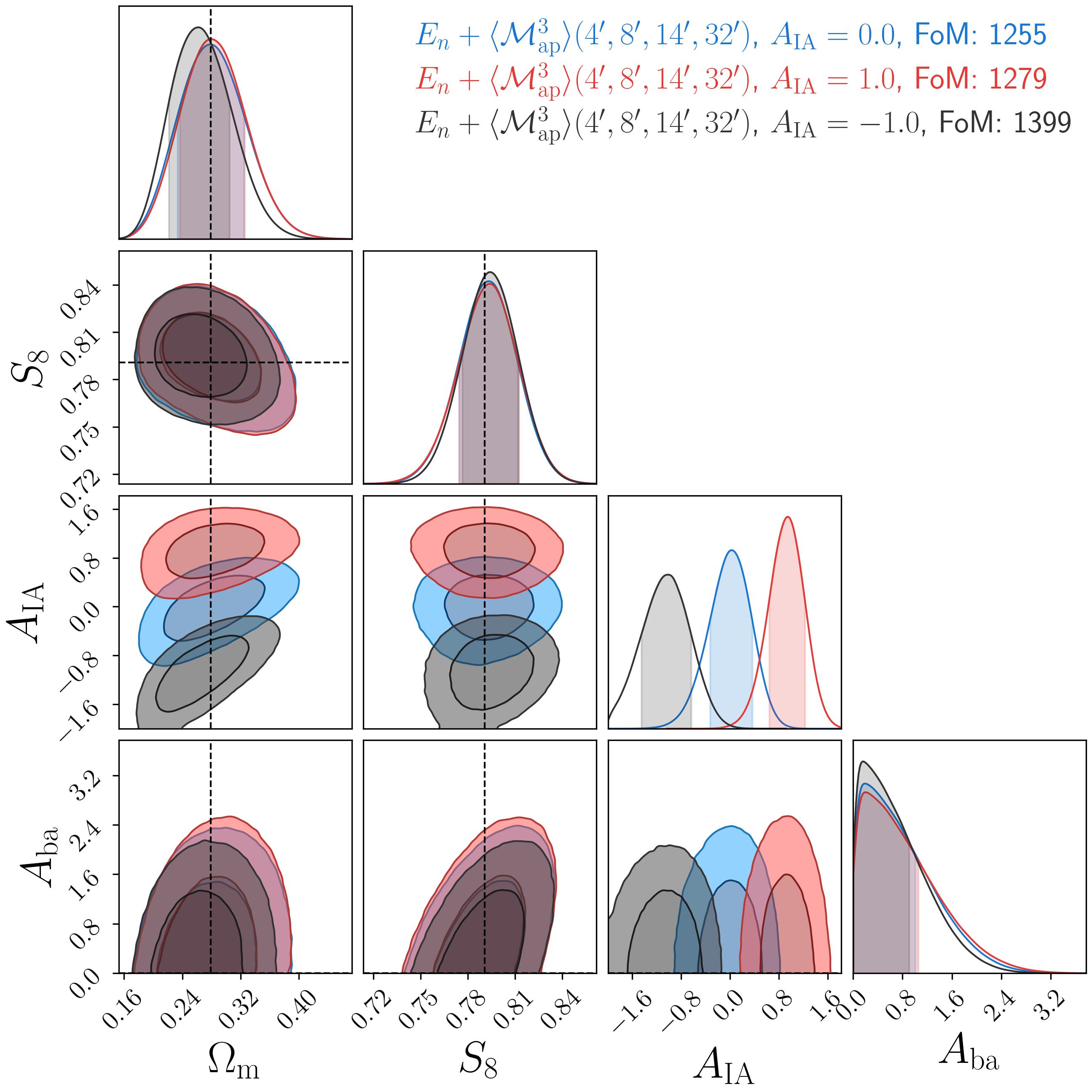}
\caption{Posterior distribution for data vectors and covariance measured from the 1944 \citetalias{Takahashi2017} noisy $g$ mock data. Here the \citetalias{Takahashi2017} data vector is infused with IA measured from the cosmo-SLICS+IA. The modelling is described in Sect.~\ref{sec:IA_modelling}.}
\label{fig:MCMC^{(i)}A}
\end{figure}

\begin{figure*}[ht]
\includegraphics[width=\textwidth]{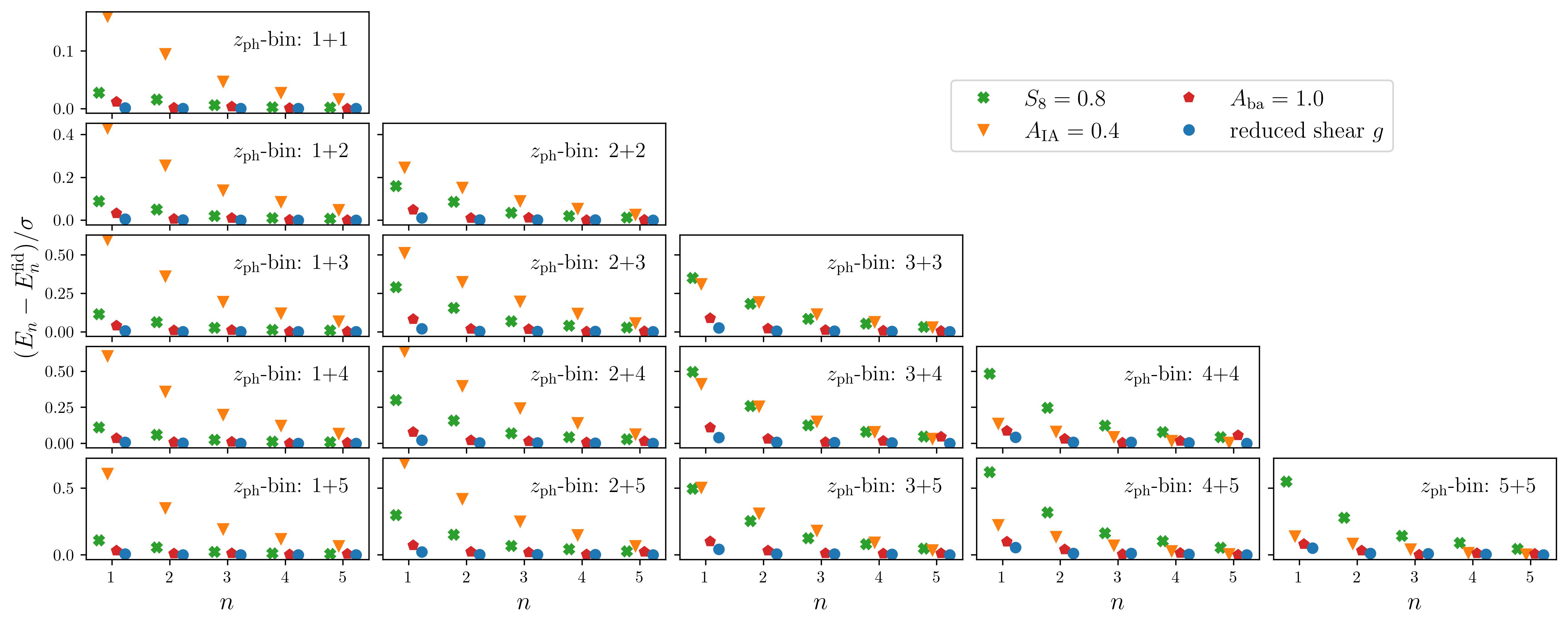}
\caption{Illustration of several effects that change the $E_n$ model vector. The fiducial model $E_n^\mathrm{fid}$ is without reduced shear, $A_\mathrm{IA}=0$, $A_\mathrm{ba}=0$, $S_8=0.78$ and $\Omm=0.25$. For the different points, one of the effects is changed. We scaled the model differences by the KiDS-1000 uncertainty, indicating the relevance of the change in the context of this work.}
\label{fig:Cosebis_Emodes_modelchecks}
\end{figure*}

\begin{figure*}[ht]
\includegraphics[width=\textwidth]{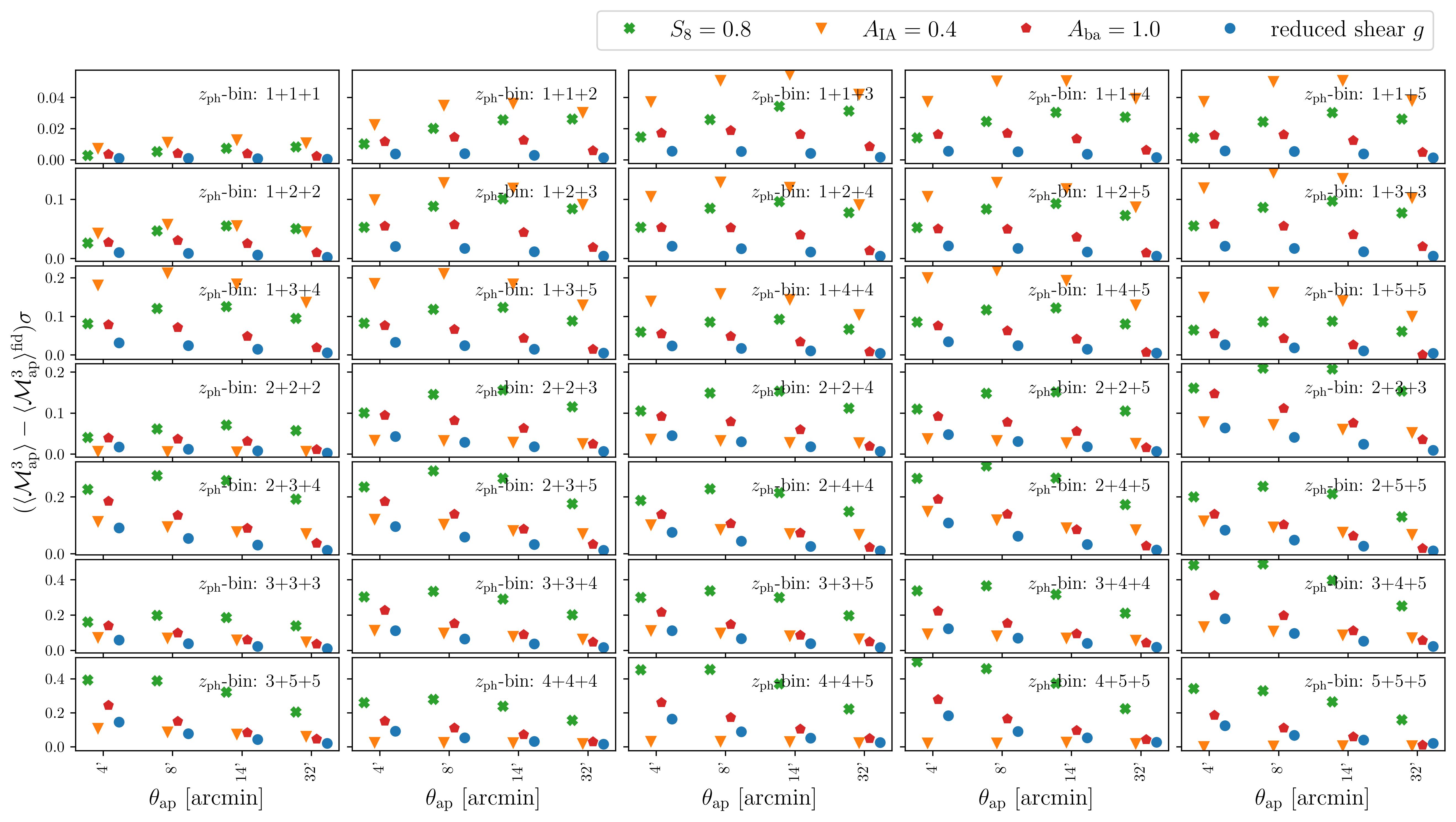}
\caption{Same as Fig.~\ref{fig:Cosebis_Emodes_modelchecks} but for the $\MapMapMap$ model vector.}
\label{fig:Map3_equalscale_modelchecks}
\end{figure*}

As a further modelling check, we investigate in Fig.~\ref{fig:Cosebis_Emodes_modelchecks} and Fig.~\ref{fig:Map3_equalscale_modelchecks} the impact on our model vector of several ingredients, namely the reduced shear correction, the infusion of baryonic feedback processes and intrinsic alignments (IA). As a reference, we always use a model without reduced shear, $A_\mathrm{IA}=0$, $A_\mathrm{ba}=0$, $S_8=0.78$, and $\Omm=0.25$. We then only change one of these effects while fixing the others. We also show the impact on the model vector when we change the cosmology to $S_8=0.8$. The change of the model vectors is divided by the square root of the covariance diagonal, indicating the relevant importance given the KiDS-1000 uncertainty. 
The reduced shear and baryon feedback correction are always subdominant compared to the impact of $S_8$ or IA. For the higher $z_\mathrm{ph}$-bins, the $S_8$ change dominates, and the IA dominates if $z_\mathrm{ph}$-bin one or two are included. This result is unsurprising given that the lower $z_\mathrm{ph}$-bins are more affected by IA as the overall shear signal is much lower. However, we observe especially for the $\MapMapMap$ that the largest IA S/N emerge if lower $z_\mathrm{ph}$-bins are combined with large $z_\mathrm{ph}$-bins, which is due to the fact that the large $z_\mathrm{ph}$-bins drastically deplete the noise. The takeaway from this investigation is that for the current KiDS-1000 analysis, our model correction for the reduced shear and baryon feedback is sufficient. The biggest impact, especially if lower $z_\mathrm{ph}$-bins are included, comes from IA, which is the biggest concern of current weak lensing analyses. 

\section{Additional material}

\begin{figure*}[ht]
\includegraphics[width=\textwidth]{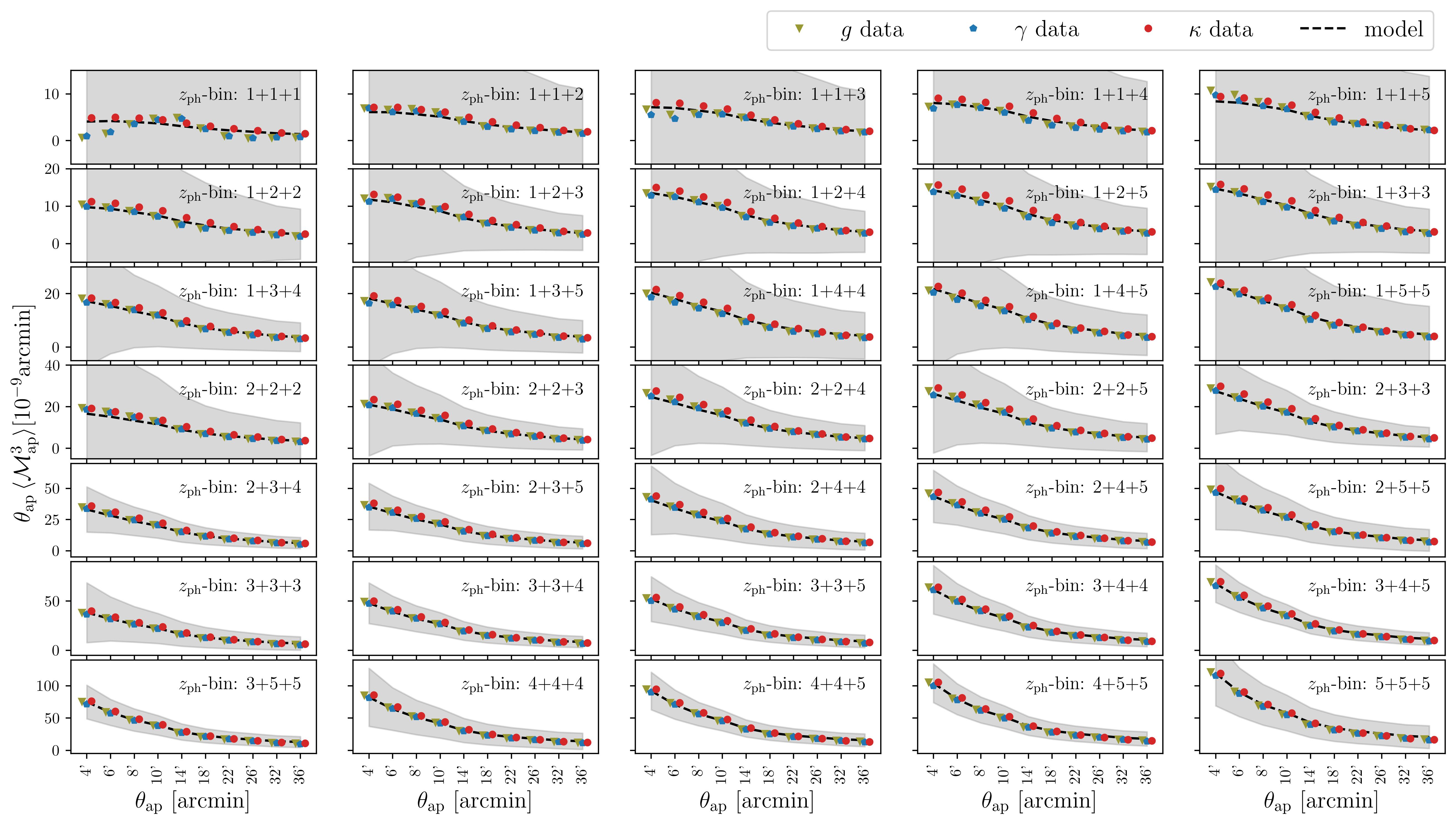}
\caption{Same as Fig.~\ref{fig:Map3_equalsclae_selected}, but here the measured data and model are the $\MapMapMap$ vector for equal-scale aperture filter radii $\theta_\mathrm{ap}\in\{\ang{;4}, \ang{;6}, \ang{;8}, \ang{;10},\ang{;14}, \ang{;18},\ang{;22}, \ang{;26}, \ang{;32}, \ang{;36}\}$ in the \citetalias{Takahashi2017} mock data for $z_\mathrm{ph}$-bin combinations. 
}
\label{fig:Map3_equalsclae}
\end{figure*}

\begin{figure*}
\includegraphics[width=\textwidth]{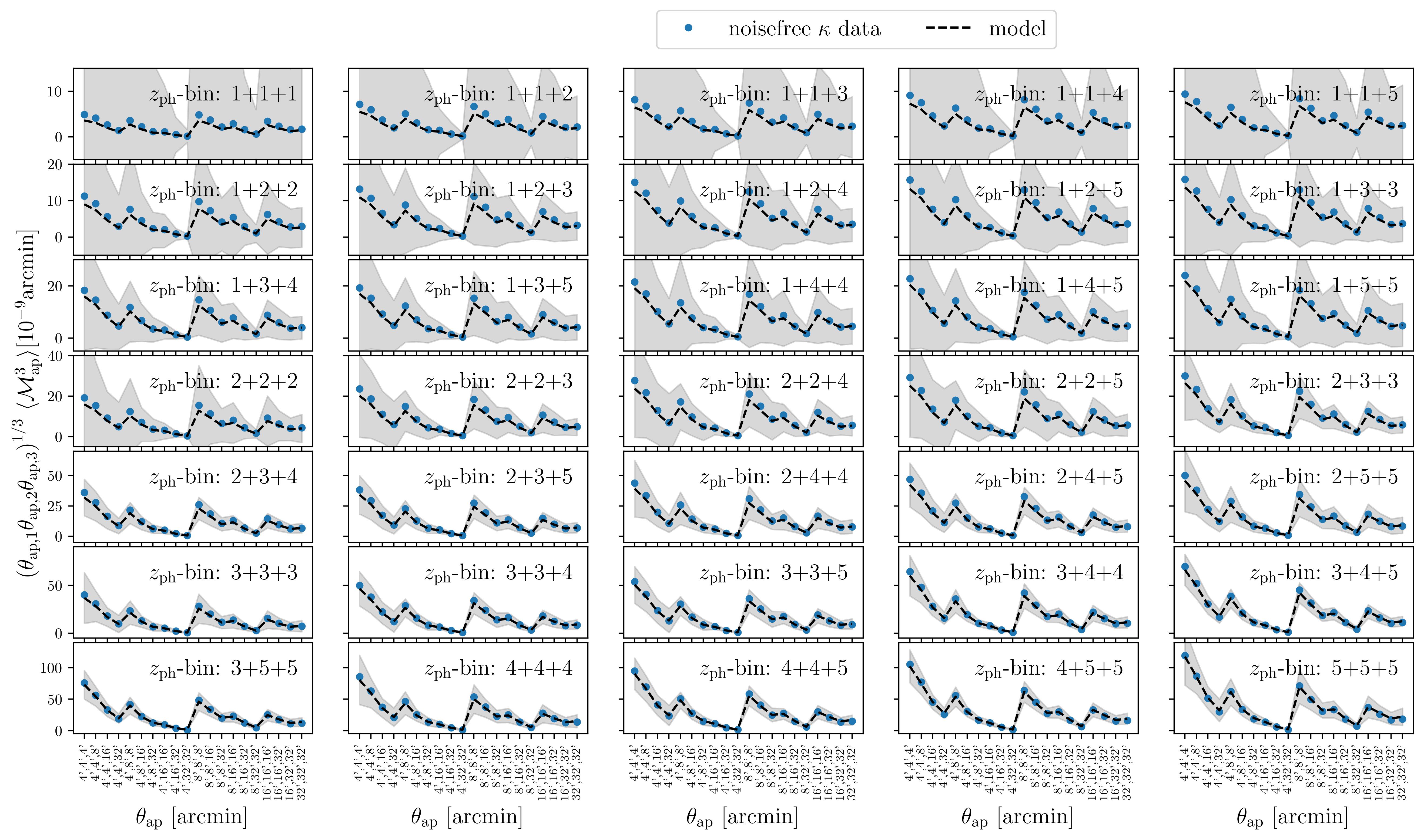}
\caption{Measured data and model $\MapMapMap$ vector for all combinations of aperture filter radii $\theta_\mathrm{ap}\in\{\ang{;4}, \ang{;8}, \ang{;16}, \ang{;32}\}$. The data vector results from one full-sky convergence realisation without shape noise. The grey band shows the expected uncertainty of KiDS-1000 estimated with the convergence maps. The $\MapMapMap$ is scaled by the third root of the product of the corresponding filter radii.}
\label{fig:Map3_allsclae}
\end{figure*}

This section offers a collection of figures to support the analysis in the main text. In Fig.~\ref{fig:Map3_equalsclae}, we show the full $\MapMapMap$ data vector for equal filter radii, from which we showed a subpart in Fig.~\ref{fig:Map3_equalsclae_selected}. 
In Fig.~\ref{fig:Map3_allsclae}, we show the $\MapMapMap$ data vector if non-equal filter radii are used and measured from the convergence \citetalias{Takahashi2017} maps. 

\end{appendix}
%
%

%

\end{document}